\documentstyle [12pt,twoside]{article}
\def\baselinestretch{1.0}  
\begin{document}
\par\noindent
to appear in {\it Physical Review E}
\vfill\small
\centerline{\huge\bf Geometric Interpretation of}
\vskip .2in
\centerline{\huge\bf Chaos in Two-Dimensional}
\vskip .24in
\centerline{\huge\bf  Hamiltonian Systems}
\vskip .3in
\centerline{\large Henry E. Kandrup\footnote{\normalsize
Electronic address: kandrup@astro.ufl.edu}}
\vskip .1in
\centerline{\it Department of Astronomy and Department of Physics and}
\vskip .05in
\centerline{\it Institute for Fundamental Theory, University of Florida, 
Gainesville, FL 32611}
\vskip .25in
This paper exploits the fact that Hamiltonian flows associated with a 
time-independent $H$ 
can be viewed as geodesic flows in a curved
manifold, so that the problem of stability and the onset of chaos hinge on
properties of the 
curvature $K_{ab}$ entering into the Jacobi 
equation. 
Attention focuses on ensembles of orbit segments evolved in representative 
two-dimensional potentials, examining how such 
properties as orbit type, 
values of short time Lyapunov exponents ${\chi}$, complexities of 
Fourier spectra, and locations of initial conditions on a 
surface of section correlate with the mean value and dispersion, 
${\langle}{\tilde K}{\rangle}$ and ${\sigma}_{\tilde K}$, of the (suitably 
rescaled) trace of $K_{ab}$. Most analyses of chaos in this context
have explored the effects of negative curvature, which implies a divergence of 
nearby trajectories. The aim here is to 
exploit instead a point stressed recently by Pettini (Phys. Rev. {\bf E 47},
828 [1993]), namely that 
geodesics can be chaotic even if $K$ is everywhere positive, 
chaos in this case arising as 
a parameteric instability 
triggered by regular variations in $K$ along the orbit. For ensembles of 
fixed energy, containing both regular and chaotic segments, simple patterns 
exist connecting ${\langle}{\tilde K}{\rangle}$ and 
${\sigma}_{\tilde K}$ for different segments both with each other and with 
the short time ${\chi}$: Often, but not always, there is a nearly one-to-one 
correlation between 
${\langle}{\tilde K}{\rangle}$ and ${\sigma}_{\tilde K}$, a plot of these two
quantities approximating a simple curve. Overall ${\chi}$ varies smoothly
along the curve, certain regions corresponding to regular and ``confined'' 
chaotic orbits where ${\chi}$ is especially small. Chaotic segments located
furthest from the regular regions tend systematically to have the largest
${\chi}$'s. The values of ${\langle}{\tilde K}{\rangle}$ and
${\sigma}_{\tilde K}$ (and in some cases ${\chi}$) for regular orbits
also vary smoothly as a function of the ``distance'' from the chaotic phase
space regions, as probed, e.g., by the location of the initial condition on a
surface of section. Many of these observed properties can be understood 
qualitatively in terms of a one-dimensional Mathieu equation, in which 
parametric instability is introduced in the simplest possible way.
\vskip .2in
\par\noindent
PACS number(s): 05.45.+b, 03.20.+i, 02.40.+m
\vfill
\par\noindent
\eject
\centerline{\bf I. INTRODUCTION AND MOTIVATION}
\vskip .1in
It is well known [1] 
that the flow associated with a time-independent Hamiltonian
$H={1\over 2}{\delta}^{ab}p_{a}p_{b}+ V(x^{a})$
can be reformulated as a geodesic flow in a curved, but conformally flat,
manifold. Specifically, let 
$$ds^{2}=W(x^{a}){\delta}_{ab}dx^{a}dx^{a}, \eqno(1)$$
where $E$ is the conserved energy associated with the time-independent $H$
and the conformal factor
$$W(x^{a})=E-V(x^{a}) \eqno(2)$$
is equal numerically to the kinetic energy associated with a trajectory at
the point $x^{a}$. It then follows that, with the further identification
$$ds=\sqrt{2}\;W\,dt, \eqno(3) $$
the geodesic equation for motion in the metric $g_{ab}=W{\delta}_{ab}$ 
is completely equivalent to the Hamilton equations
$${dx^{a}\over dt}={{\partial}H\over {\partial}p_{a}} \qquad {\rm and} \qquad
{dp_{a}\over dt}=-{{\partial}H\over {\partial}x^{a}}. \eqno(4) $$

This implies that the confluence or diverence of nearby trajectories $x^{a}(s)$
and $[x+{\xi}]^{a}(s)$ is determined by the Jacobi equation, i.e., the 
equation of geodesic deviation, which takes the form 
$${D^{2}{\xi}^{a}\over Ds^{2}}=-R^{a}_{\;bcd}u^{b}u^{d}{\xi}^{c}{\;}{\equiv}
-K^{a}_{\;c}\,{\xi}^{c} , \eqno(5) $$
where $R_{abcd}$ is the Riemann tensor associated with $g_{ab}$ and
$D/Ds=u^{a}{\nabla}_{a}$ denotes a directional derivative along 
$u^{a}=dx^{a}/ds$. Linear stability or lack thereof for the trajectory 
$x^{a}(s)$ is thus related to $R_{abcd}$ or, more precisely, to the 
curvature $K^{a}_{\;c}$. If, e.g., $R_{abcd}$ is everywhere 
negative, so that $K^{a}_{\;c}$ always has one or more negative eigenvalues, 
the trajectory must be linearly unstable. 

It would seem intuitive that, if the curvature of $g_{ab}$ is everywhere 
negative, so that nearby trajectories always tend to diverge, 
every geodesic will behave in a fashion that is manifestly chaotic. If one 
assumes that the manifold is compact, so that trajectories are
restricted to a region of finite volume, this intuition can be elevated to 
a theorem. For example, geodesic flows on a compact manifold with constant
negative curvature are necessarily $K$-flows, where  generic 
ensembles of initial conditions evolve towards a microcanonical 
distribution at a rate set by the magnitude of the curvature [2].
If the curvature is everywhere negative but not constant, the flow is more
complex, but one can still infer 
[3] a chaotic evolution towards a
microcanonical distribution at a rate bounded from below by the least negative
value of the curvature.

When the curvature is not everywhere negative, much less is known.
Nevertheless, the preceeding has motivated the expectation that, 
in many dynamical systems, chaos should be associated with (regions of)
negative curvature. In particular, several authors (cf.~[4,5])
have sought to use negative curvature to explain the fact 
that the gravitational $N$-body problem for a large number of 
objects of comparable mass is chaotic in the sense that the evolution 
manifests an exponentially sensitive dependence on initial conditions.

However, as stressed recently by Pettini [6], not all chaos can be associated 
with negative curvature. In particular, one can have large measures of chaotic 
orbits even for systems and energies where $K_{ab}$ is everywhere positive.
In retrospect, this is easy to understand. Viewing the Jacobi equation
as a matrix equation, one can solve at any given point in space to derive
eigenvectors $\{ {\bf X}_{i} \}$ and eigenvalues $\{ {\lambda}_{i} \}$,
each pair solving a linear equation of the form
$${D^{2}{\bf X}_{i}\over Ds^{2}}=-{\lambda}_{i}{\bf X}_{i}. \eqno(6) $$
When the curvature is everywhere positive, ${\lambda}_{i}{\;}{\ge}{\;}0$, so 
that
the solutions are oscillatory rather than exponential. If the ${\lambda}_i$'s 
were constant along the trajectory, one could thus infer stable oscillations.
In general, however, the ${\lambda}_{i}$'s are not constant, depending instead 
on the unperturbed $x^{i}(s)$ since $R_{abcd}$ and $u^{a}$ both change along
the trajectory. It follows that, even assuming an everywhere positive 
curvature, eq. (6) should be interpreted as an oscillator equation
$${D^{2}{\bf X}_{i}\over Ds^{2}}=-{\Omega}^{2}_{i}(s){\bf X}_{i}. \eqno(7) $$
The obvious point is that solutions to eq. (7) can manifest a parametric
instability.

Because the coordinates and velocity of a regular orbit are periodic, the
frequency ${\Omega}(s)$ must also be periodic, so that 
eq. (7) reduces to an oscillator equation with periodic modulation, e.g., a 
Hill equation. If the unperturbed geodesic is to be stable, so that it can 
exist as a regular orbit, it must be that solutions to this equation represent 
bounded oscillations, a condition that implies nontrivial restrictions on the 
time-dependence of ${\Omega}$. If these restrictions are not satisfied, the
geodesic must instead correspond to a chaotic orbit. 

As observed by Cerruti-Sola and Pettini [7], this intuition is particularly 
simple to implement for two-dimensional systems. In this case, 
$K^{a}_{\;b}$ corresponds to a symmetric 
$2\times 2$ matrix, one eigenvalue of which is necessarily zero, this 
corresponding to neutral stability with respect to infinitesimal translations
along the orbit from $x^{a}(s)$ to $x^{a}(s+{\delta}s)$. It follows that, in 
interpreting the behaviour of a small perturbation, it suffices to restrict 
attention to perturbations in the single direction orthogonal to
the velocity $u^{a}$, so that one is reduced to a single scalar equation.

More explicitly, by exploiting the fact that, in two dimensions, the Riemann
tensor has only one independent nonzero component, say $R^{x}_{\;yxy}$, it is
easily seen that
$$K^{a}_{\;b}=
\left(\matrix{R^{x}_{\;yxy}(u^{y})^{2} & -R^{x}_{\;yxy}u^{x}u^{y} \cr
		-R^{x}_{\;yxy}u^{x}u^{y} & R^{x}_{\;yxy}(u^{x})^{2}\cr}\right),
\eqno(8)$$
whence follows that there are two eigenvalues, namely
${\lambda}=0$ and 
${\lambda}=R^{x}_{\;yxy}[(u^{x})^{2}+(u^{y})^{2}]=K^{x}_{\;x}+K^{y}_{\;y}$.
Transforming from $u^{a}$ to the physical momentum $p_{a}$ and recalling that
$W$ is equal numerically to the kinetic energy, it follows that the 
nonzero eigenvalue
$${\lambda}=R_{xyxy}/W^{2}, \eqno(9) $$
where, explicitly,
$$R_{xyxy}={1\over 2}{\Biggl[}{{\partial}^{2}V\over {\partial}x^{2}}+
{{\partial}^{2}V\over {\partial}y^{2}}{\Biggr]}+
{1\over 2W}{\Biggr[}{\biggl(}{{\partial}V\over {\partial}x}{\biggr)}^{2}+
{\biggl(}{{\partial}V\over {\partial}y}{\biggr)}^{2}{\Biggr]}. \eqno(10) $$
The component of ${\xi}^{a}$ orthogonal to $u^{a}$ thus satisfies 
$${d^{2}{\xi}_{1}\over ds^{2}}=-K{\xi}_{1}=
-{1\over 2}\,R{\xi}_{1} , \eqno(11) $$
where
$K{\;}{\equiv}{\;}K_{x}^{\;x}+K_{y}^{\;y}$
and $R$ denotes the scalar curvature. Alternatively, translating back into 
physical time $t$ one finds (cf. [7])
$${d^{2}{\xi}_{1}\over dt^{2}}-{1\over W}{dW\over dt}{d{\xi}_{1}\over dt}=-
2R_{xyxy}{\xi}_{1}=-W^{2}K{\xi}_{1}. \eqno(12) $$

Cerruti-Sola and Pettini [7] have studied representative orbits in one 
prototypical two-dimensional potential, namely the H\'enon-Heiles potential 
[8], demonstrating thereby that one can effect a translation between 
various orbital properties as viewed in the ordinary Hamiltonian language and 
as viewed in this geometric language. However, it is also useful to study the 
statistical properties of ensembles of orbit segments since this facilitates 
a search for bulk 
regularities connecting different properties of representative orbits. Thus, 
in particular, such an investigation can provide important information about 
how quantities like short time Lyapunov exponents ${\chi}(t)$ [9]
correlate with properties of the curvature $K_{ab}$ as evaluated along 
some orbit. This is of interest for chaotic orbits, where one knows that 
changing values of short time Lyapunov exponents can reflect phase 
space transport through cantori [10] and other topological obstructions 
[11] 
and/or the overall complexity of an orbit segment, as probed by its Fourier 
spectrum [12]. 
This is also useful for regular orbits where, for different initial 
conditions, the short time exponent ${\chi}(t)$ can converge towards the 
asymptotic Lyapunov exponent ${\chi}_{\infty}={\chi}(t\to\infty)$ at very 
different rates.
Indeed, for fixed time $t$ the value of ${\chi}(t)$ for different regular 
orbits with the same energy $E$ can vary by more than an order of magnitude.

Equations (11) and (12) might suggest that the natural quantity upon which 
to focus is $K$ or $W^{2}K$. However, when realised as a function of $W$ and
its derivatives, both these quantities involve (cf. eq. [10]) division by 
positive powers
of $W$, which, for small values of $W$, can lead to skewed statistics for
a finite sampling. For this reason, it was discovered that cleaner 
results were obtained by focusing on 
$${\tilde K}{\;}{\equiv}{\;}W^{3}K=WR_{xyxy} . \eqno(13) $$
Physically this combination arises if one introduces a new time coordinate
${\tau}$ satisfying $ds=W^{3/2}d{\tau}$, so that
$${d^{2}{\xi}_{1}\over d{\tau}^{2}}-{3\over 2W}{dW\over d{\tau}}
{d{\xi}_{1}\over d{\tau}}=-W^{3}K{\xi}_{1}. \eqno(14) $$
The fact that the statistical properties of ${\tilde K}=W^{3}K$ correlate 
very strongly with properties of the unperturbed orbit suggests that
it is the coefficient of ${\xi}_{1}$, rather than its time derivative, that
is responsible for much of the orbit's observed behaviour.

The work described in this paper involved examining the quantity ${\tilde K}$
as evaluated along different orbit segments, probing in particular values of 
the mean ${\langle}{\tilde K}{\rangle}$ and the dispersion 
${\sigma}_{\tilde K}$. The resulting data were used to establish trends 
related to these quantities, including how different segments -- both regular 
and chaotic -- fit into the 
${\langle}{\tilde K}{\rangle}\,$-$\,{\sigma}_{\tilde K}$ 
plane, and how the values assumed by these quantities depend on other orbital 
characteristics, e.g., on whether a regular orbit is a box or a loop or 
whether a chaotic segment looks nearly regular or particularly complex. 
Especially striking were correlations between
${\langle}{\tilde K}{\rangle}$ or ${\sigma}_{\tilde K}$ and the values of 
short time Lyapunov exponents ${\chi}(t)$ computed for the same segments. 
Given that the value of ${\chi}$ is strongly correlated with the overall 
complexity of the orbit, as probed by its Fourier spectrum [12], such 
correlations also connect statistical properties of ${\tilde K}$ with 
the shape of the orbit, as viewed in configuration space.

As noted by Pettini (private communication), the preceding justification
for focusing ${\tilde K}=W^{3}K$, rather than $W^{2}K$, is potentially suspect 
since, at least in principle, any nontrivial time reparameterization can
significantly alter the stability properties of geodesics. Fortunately,
however, there is another interpretation which may perhaps be more easily 
justified: Computing the average of $W^{2}K$ along a geodesic can be
interpreted as involving a ratio of integrals 
$\int\,\sqrt{g}\,dxdy W^{2}K\,/\,\int\,\sqrt{g}\,dxdy , $
where $g$ denotes the determinant of the metric $g_{ij}$ and the integration
extends over the regions of the manifold along which the geodesic moves.
However, it follows from eq. (2) that this reduces to
$\int\,dxdy W^{3}K\,/\,\int\,dxdy\, W, $
and, to the extent that $\int\,dxdy\,W$ is approximately constant (as a 
consequence of
virialisation), one is effectively averaging the quantity $W^{3}K$.

The results presented below derive from an analysis of orbit segments in
two different representative two-dimensional potentials. The first of these,
$$V(x,y)=-(x^{2}+y^{2})+{1\over 4}
(x^{2}+y^{2})^{2}-{1\over 4}x^{2}y^{2} , \eqno(15) $$
corresponds to the so-called dihedral potential of Armbruster {\it et al}
[13], for one particular set of parameter values. The second, 
$$V(x,y)={1\over 2}{\Bigl(}x^{2}+y^{2}{\Bigr)}+x^{2}y-{1\over 3}y^{3}
+{1\over 2}x^{4}+x^{2}y^{2}+{1\over 2}y^{4}$$
$$+x^{4}y+{2\over 3}x^{2}y^{3}-{1\over 3}y^{5}
+{1\over 5}x^{6}+x^{4}y^{2}+{1\over 3}x^{2}y^{4}+{11\over 45}y^{6},
\eqno(16) $$
represents the sixth order truncation of the Toda [14] lattice potential
(recall that the H\'enon-Heiles potential can be derived as the third order
truncation of the Toda potential). These are very different
qualitatively but, nevertheless, much of the observed behaviour is very
similar for orbit segments in both potentials. In the truncated Toda potential,
${\tilde K}$ is always non-negative. However, for certain energies in the 
dihedral potential ${\tilde K}$ can be negative along parts of some orbits, 
although ${\tilde K}$ tends to be positive most ($>75 - 80\%$) of the time. 

Ensembles of orbit segments with fixed energy $E$ were generated by sampling
the $E$= constant hypersurface and then evolving typically for a total time 
$t=256$, this in units where a typical crossing time (i.e., the time
required for an orbit to cross from one side of the potential to the other)
$t_{cr}{\;}{\sim}{\;}1-2$. This is a reasonable time interval to consider
because chaotic segments in these potentials tend to exhibit significant
qualitative changes on a time scale ${\sim}{\;}100-200t_{cr}$ [15]. However, 
it was verified that similar results obtain for somewhat longer and shorter 
total times. In most cases, the initial ensemble was generated by setting 
$x=0$, uniformly sampling the energetically allowed portions of the 
$y\,$-$\,p_{y}$ plane, and then computing an initial $p_{x}>0$ as a function 
of $x$, $y$, $p_{y}$, and $E$. The orbits were integrated using a fourth order
Runge-Kutta scheme with a time step ${\delta}t=10^{-4}$.

Strictly speaking, when extracting statistical properties of ${\tilde K}$ it 
is most 
natural to analyse a time series which records relevant quantities at fixed
intervals of geodesic time $s$, rather than at fixed intervals of ``physical''
time $t$, which is achieved most easily by solving the geodesic equation 
associated with $W{\delta}_{ab}$ rather than the original Hamilton equations. 
However, it was found that the basic correlations involving quantities like 
${\langle}{\tilde K}{\rangle}$ and ${\sigma}_{\tilde K}$ were equally apparent 
for both sorts of time series. The discussion here focuses primarily on data 
recorded at fixed intervals ${\delta}t$. This has the advantage that the 
conclusions derived here for the dihedral and truncated Toda potentials
can be easily tested for other potentials, without the inconvenience of 
explicitly reformulating the evolution as a geodesic flow. 

Section II describes various trends and correlations observed in the numerical
experiments, demonstrating in particular the existence of striking regularities
connecting quantities like ${\langle}{\tilde K}{\rangle}$, 
${\sigma}_{\tilde K}$, and ${\chi}$, many of which can be interpreted in terms 
of other physical properties of the orbits. Section III summarises the 
principal conclusions and then shows that, not surprisingly, many of these can 
in fact be understood in terms of a simple one-dimensional Mathieu equation.
\vskip .2in
\centerline{\bf II. OBSERVED CORRELATIONS AMONGST 
${\langle}{\tilde K}{\rangle}$, ${\sigma}_{\tilde K}$, AND ${\chi}$}
\vskip .1in
\centerline{\bf A. Correlations between ${\langle}{\tilde K}{\rangle}$ and 
${\sigma}_{\tilde K}$}
\vskip .1in
In most, albeit not all, cases, i.e., for most energies in both potentials,
the values of the mean ${\langle}{\tilde K}{\rangle}$ and the dispersion 
${\sigma}_{\tilde K}$ of different orbit segments are strongly correlated. 
Rather than filling a large portion of the 
${\langle}{\tilde K}{\rangle}\,$-$\,{\sigma}_{\tilde K}$ 
plane, the orbit segments tend to fall, at least approximately, along a single 
curve. For the dihedral potential, this curve is typically quite thin; for the 
truncated Toda potential, it can be significantly thicker. Moreover, 
this curve is typically characterised overall by a negative 
slope, so that orbits with larger ${\langle}{\tilde K}{\rangle}$ have smaller 
${\sigma}_{\tilde K}$.

Figures 1 and 2 exhibit plots of ${\sigma}_{\tilde K}$ as a function of 
${\langle}{\tilde K}{\rangle}$ for several different energies in, respectively,
the dihedral and truncated Toda potentials. Figures  1d - h and Figs.
2c and d, each characterised by a single curve, perhaps with one or more
intermittent gaps (as in Figs.~1e and f) or a slight ``wiggle'' (as in Fig.~1h)
represent typical behaviour. The fact, manifest visually, that the values of 
${\langle}{\tilde K}{\rangle}$ and ${\sigma}_{\tilde K}$ are strongly 
correlated can be quantified by computing the rank correlation 
${\cal R}({\langle}{\tilde K}{\rangle},-{\sigma}_{\tilde K})$. For these 
typical cases in the dihedral potential, the correlation
between ${\langle}{\tilde K}{\rangle}$  and $-{\sigma}_{\tilde K}$ is 
usually very strong, ${\cal R}>0.98$ or more. For the truncated Toda 
potential, the correlation is often somewhat weaker,
${\cal R}({\langle}{\tilde K}{\rangle},-{\sigma}_{\tilde K}){\;}{\sim}{\;}
0.9-0.95$, but still significant.

This sort of correlation between ${\langle}{\tilde K}{\rangle}$ and
${\sigma}_{\tilde K}$, seemingly suggestive of ordered behaviour, is perhaps 
not surprising for regular orbits, where the motion is multiply periodic. 
Thus, e.g., it is easy to envision a sequence of regular box or loop orbits
characterised by smoothly varying values of ${\langle}{\tilde K}{\rangle}$
or ${\sigma}_{\tilde K}$. However, such correlations might seem less expected 
for chaotic orbits where the motion is aperiodic. It is therefore significant 
that, when viewed in such 
a ${\langle}{\tilde K}{\rangle}\,$-$\,{\sigma}_{\tilde K}$ plot,
regular orbits need not stand out in any obvious way. Consider, e.g.,
the orbit ensemble with $E=4.0$ used to generate Fig.~1f. Here one finds
that there are three distinct types of regular orbits, 
a large number of orbits with $59<{\langle}{\tilde K}{\rangle}<63$ and 
$9<{\sigma}_{\tilde K}<26$, 
a large number of orbits with $54<{\langle}{\tilde K}{\rangle}<57$ and
$33<{\sigma}_{\tilde K}<36$, and a small number of orbits with
$43<{\langle}{\tilde K}{\rangle}<45$ and $49<{\sigma}_{\tilde K}<50$.
It is clear that all the points below the gap in Fig.~1f are associated
with regular orbits, and a careful examination of the data points allows one
to distinguish minute differences between the locations of the regular and
chaotic orbits in the intermediate regime, 
$54<{\langle}{\tilde K}{\rangle}<57$ and $33<{\sigma}_{\tilde K}<36$. However,
it is evident that, overall, the regular and chaotic orbits coexist along a 
relatively narrow curve [16].

This can be associated tentatively with the fact that, even though chaotic 
orbits are intrinsically aperiodic, finite chaotic segments often manifest a 
fair amount of regularity. Thus, e.g., as discussed more carefully below,
one observes oftentimes that the power 
spectra, $|x({\omega})|$ and $|y({\omega})|$, for a chaotic segment typically 
appear visually to be constructed from ``pieces'' appropriate for a small 
number of regular orbits [12,17]. This interpretation is especially natural 
given the recognition that the most striking exception
to the simple pattern described hitherto is associated with very low energies
in the dihedral potential, where few, if any, regular orbits exist. 

Figure 1a, a seemingly structureless set of points completely different from
the lines observed in the remaining panels of Figs. 1 and 2, derives from an 
ensemble of segments with $E=-0.05$, an energy where a sampling of nearly 
$1000$ different initial conditions yielded only chaotic orbits. At slightly
higher energies, regular orbits begin to appear and, unlike the chaotic
orbits, they seem concentrated largely along lines in the 
${\langle}{\tilde K}{\rangle}\,$-$\,{\sigma}_{\tilde K}$ plane. Thus, e.g., as 
illustrated in Fig. 1b, for $E=+0.05$ there are four different types of 
regular orbits, three of these associated with the three conspicuous lines and 
the fourth associated with the largest values of ${\sigma}_{\tilde K}$ above 
the central chaotic region. As the energy is raised to yet higher values, the 
regular families associated with the two upper lines eventually disappear, so 
that, e.g., as illustrated in Fig. 1d for $E=1.0$, all the orbits fit 
approximately onto a single curve, with the chaotic orbits at high 
${\sigma}_{\tilde K}$ and the regular orbits at low ${\sigma}_{\tilde K}$.
When the energy is raised to a higher value $E{\;}{\sim}{\;}3$, other regular
orbit families appear, leading eventually to the aforementioned behaviour at
$E=4.0$. 

The observed behaviour at low energies in the truncated Toda potential is
quite different, this presumably reflecting the fact that, in this case, at
very low energies there is no global stochasticity. For energies $E<5.0$ or so,
all the orbits present, both regular and chaotic, fit into two distinct lines
in the ${\langle}{\tilde K}{\rangle}\,$-$\,{\sigma}_{\tilde K}$ plane. 
Consider,
e.g., the energy $E=0.5$ exhibited in Fig. 1a. Here most of the upper line is 
occupied by loop orbits which manifest a discrete $2{\pi}/3$ rotation
symmetry, whereas the lower line is occupied completely by box and
banana orbits that break this symmetry. There are only a
very few chaotic orbits at this energy, and all of them fit at the high
${\sigma}_{\tilde K}$ end of the upper line. As the energy increases, the 
lower line eventually shrinks and merges into the upper line, so that,
ultimately, different families of regular orbits coexist with chaotic orbits
along a single thickened curve.
\vskip .2in
\centerline{\bf B. Correlations between curvature and ${\chi}(t)$}
\vskip .1in
\par
Ordinary Lyapunov exponents ${\chi}$, which [18] probe the average instability 
of some trajectory in an asymptotic $t\to\infty$ limit, manifest a fundamental 
distinction between regular and chaotic orbits. For regular orbits all the 
Lyapunov exponents vanish identically, whereas chaotic orbits have at least 
one ${\chi}$ that is positive. Orbits in a $D$-dimensional system have $2D$ 
distinct Lyapunov exponents, these corresponding to 
perturbations in $2D$ independent phase space directions. For a 
time-independent Hamiltonian system, two of these exponents must vanish (this 
reflecting neutral stability with respect to perturbations that translate an 
orbit from $x^{a}(t)$ to $x^{a}(t+{\delta}t)$ and to perturbations orthogonal 
to the constant energy surface) and the remaining exponents must come in pairs,
${\pm}{\chi}$. It follows that, for two-dimensional Hamiltonian systems, the 
only fundamental distinction is between regular orbits, for which all the
${\chi}$'s vanish, and chaotic orbits, which have one exponent ${\chi}>0$.

However, one can also introduce short time Lyapunov exponents ${\chi}(t)$ which
provide information about the average instability of orbit segments over 
a finite interval. Thus, in particular, for any phase space perturbation
${\delta}z$, one can define [9]
$${\chi}(t){\;}
{\equiv}{\;}\lim_{t\to\infty}\lim_{{\delta}Z(0)\to 0}{1\over t}\;
\ln {\Biggl[}{||{\delta}Z(t)||\over ||{\delta}Z(0)||}{\Biggr]}, \eqno(17) $$
where $||\,.\,||$ represents a suitable norm. For a generic initial 
perturbation, this ${\chi}(t)$ will converge towards the largest Lyapunov
exponent in the limit $t\to\infty$, independent of the detailed choice of norm.
By contrast, at finite times the computed ${\chi}(t)$ will depend on both 
the initial perturbation and the choice of norm. Suppose, however, that 
$||\,.\,||$ is taken as the natural $L^{2}$ phase space norm, i.e., 
$||{\delta}Z||^{2}=({\delta}x)^{2}+({\delta}y)^{2}+({\delta}p_{x})^{2}+
({\delta}p_{y})^{2}.$ In this case, one knows that 
the computed ${\chi}(t)$ will be insensitive to the detailed choice
of initial perturbation for times $t{\;}{\gg}{\;}1/{\chi}(t)$. 

It follows
that, for chaotic segments integrated in the dihedral and truncated Toda
potentials for times as long as $t=256$, the computed ${\chi}(t)$ is nearly
independent of the initial ${\delta}Z$. However, as described below the 
values of ${\chi}(t)$ computed for regular orbits can, and in certain cases do,
exhibit a significant dependence on ${\delta}Z$. For the experiments described
in this paper, ${\chi}(t)$ was computed [18] by introducing a small 
perturbation of magnitude $||{\delta}Z||=10^{-10}$, evolving both unperturbed 
and perturbed initial conditions, and periodically renormalising the 
perturbation to an amplitude $||{\delta}Z||=10^{-10}$ at intervals 
${\Delta}t=10$. Unless stated otherwise, the initial perturbation was taken as 
${\delta}Z={\delta}x=10^{-10}$.


The objective here is to show that, both for regular and chaotic orbit
segments, strong correlations exist between the value of the short time 
${\chi}(t)$ and such properties of the 
curvature as 
${\langle}{\tilde K}{\rangle}$ and ${\sigma}_{\tilde K}$. Because the 
quantities ${\langle}{\tilde K}{\rangle}$ and ${\sigma}_{\tilde K}$ are 
themselves correlated, it would seem equally reasonable to look for 
correlations between ${\chi}(t)$ and either ${\langle}{\tilde K}{\rangle}$ or 
${\sigma}_{\tilde K}$ (or any combination of these two quantities). For
specificity, most of the discussion will focus on correlations
between ${\sigma}_{\tilde K}$ and ${\chi}(t)$, although several Figures 
exhibit examples of correlations between 
${\langle}{\tilde K}{\rangle}$ and ${\chi}(t)$.
Figures 3 and 4 exhibit plots of ${\chi}$ as a function of 
${\sigma}_{\tilde K}$, generated respectively for the dihedral and 
truncated Toda potentials for the same ensembles used to generate Figs. 1 
and 2.

Consider first the case of orbit ensembles evolved in the dihedral potential.
For $E=-0.05$, where there are few if any regular orbits, a plot of ${\chi}$
as a function of ${\sigma}_{\tilde K}$ shows little obvious structure:
all that one sees is a seemingly random scattering of points at values of 
${\chi}$ well separated from ${\chi}=0$. However, as $E$ increases, one begins 
to observe the existence of regular regions, these corresponding to ranges of 
${\sigma}_{\tilde K}$ where, for some segments, ${\chi}$ assumes values 
much smaller than the values associated with chaotic segments. Thus, e.g., as 
illustrated in Fig. 3b, for $E=0.05$ one observes an extended low 
${\sigma}_{\tilde K}$ band that corresponds to one regular line
in Fig.~1b, a collection of points near ${\sigma}_{\tilde K}=6.0$ 
corresponding to the second line, another collection near 
${\sigma}_{\tilde K}=7.4$ corresponding to the third line, and a 
small number of small ${\chi}$ points near ${\sigma}_{\tilde K}=7.8$ 
corresponding to the high ${\sigma}_{\tilde K}$ points in Fig.~1b.

For $E>0.2$ or so, a simpler pattern emerges which includes only two types
of regular orbits, these concentrated at especially low and high values of 
${\sigma}_{\tilde K}$ (or, equivalently, high and low values of 
${\langle}{\tilde K}{\rangle}$). At somewhat higher energies, the high 
${\sigma}_{\tilde K}$ family disappears, only to be replaced by a new regular 
family concentrated at intermediate values of ${\sigma}_{\tilde K}$.
At yet higher energies, one sees two large regular regions, concentrated at
the largest and smallest values of ${\sigma}_{\tilde K}$, along with some 
intermediate regions with small ${\chi}$'s, these associated with nearly 
vertical lines in the ${\sigma}_{\tilde K}\,$-$\,{\chi}$ plane. Viewed in a 
surface of section, the low ${\sigma}_{\tilde K}$ regular region 
corresponds to a large island of loop orbits; the high ${\sigma}_{\tilde K}$ 
region corresponds to a large island of box orbits. The regular orbits 
associated
with the vertical lines correspond to smaller islands embedded
in the stochastic sea. This structure appears to persist to very high energies.

For these relatively high energies, $E>6.0$ or so, a plot of ${\chi}$ as a 
function of ${\sigma}_{\tilde K}$ or ${\langle}{\tilde K}{\rangle}$ exhibits 
several striking regularities. One significant point, well illustrated for 
$E=10.0$ in Fig. 5,  is that the transition from regularity to chaos observed 
at low and high values of ${\sigma}_{\tilde K}$ (or high and low values of
${\langle}{\tilde K}{\rangle}$) is relatively abrupt. Thus, e.g., the small 
${\sigma}_{\tilde K}$, large ${\langle}{\tilde K}{\rangle}$  loop orbits can 
be viewed as a sequence beginning at ${\sigma}_{\tilde K}{\;}{\approx}{\;}22$ 
that terminates at a value ${\sigma}_{\tilde K}{\;}{\approx}{\;}112$, whereas 
the large 
${\sigma}_{\tilde K}$, small ${\langle}{\tilde K}{\rangle}$ boxes can be 
viewed as a sequence extending upwards from 
${\langle}{\tilde K}{\rangle}{\;}{\approx}{\;}124$ and terminating at 
a value ${\langle}{\tilde K}{\rangle}{\;}{\approx}{\;}135$. The chaotic
segments situated near the boundary with the outer regular regions tend 
typically to be ``confined'' or ``sticky'' chaotic orbits [19] trapped near 
the regular regions by cantori [10] which, oftentimes, only escape to travel 
throughout the remainder of the stochastic
sea on a time scale $t>256$.
It is also apparent from Fig. 5 that (even away from the boundaries) chaotic 
segments with values of
${\sigma}_{\tilde K}$ furthest from the high and low ${\sigma}_{\tilde K}$ 
regular regions tend systematically to have larger ${\chi}$'s than do chaotic 
segments with values of ${\sigma}_{\tilde K}$ closer to the outer regular
regions. The net result is a curve which, viewed broadly, resembles a Greek
${\Lambda}$. 

The overall shape of the ${\sigma}_{\tilde K}-{\chi}$ curve in the regular 
region depends on the choice of the initial seed ${\delta}Z$ used in computing 
the short time ${\chi}(t)$. If, as in Figs.~5a and b, one computes ${\chi}$ 
from a seed ${\delta}Z={\delta}x$, it is apparent that the segments with 
values of ${\sigma}_{\tilde K}$ (or ${\langle}{\tilde K}{\rangle}$) closest to 
the central chaotic region, $112<{\sigma}_{\tilde K}<142$ or so, tend overall 
to have larger values of ${\chi}$ than do orbits with values of 
${\sigma}_{\tilde K}$ further from this chaotic region. However, if ${\chi}$ 
is generated instead from a seed with a significant component in one of the 
other three phase space directions, this trend is significantly diminished. 
This is illustrated in Figs. 5c and d, which exhibit ${\chi}$'s generated for
the same initial conditions from 
a seed ${\delta}Z={\delta}y$. This difference presumably reflects the fact 
that, since the initial conditions were sampled from an $x=0$ surface of 
section, a perturbation ${\delta}Z={\delta}x$ tends to be more nearly aligned 
along a direction of neutral stability, namely translation from ${\bf x}(t)$ 
to ${\bf x}(t+{\delta}t)$, than do perturbations with a nonzero ${\delta}y$.


At lower energies, where the high ${\sigma}_{\tilde K}$ region is absent or
not well developed, the right side of the ${\Lambda}$ is missing. However,
one still observes that chaotic segments with ${\sigma}_{\tilde K}$ closer to
the regular region tend to have smaller values of ${\chi}$ than do chaotic
segments far from the regular regions; and, at least for a seed 
${\delta}Z{\;}{\approx}{\;}{\delta}x$, that regular orbits closer to the
chaotic region tend to have larger values of ${\chi}$.

One obvious complication associated with the common ${\Lambda}$ pattern, well
illustrated in Fig. 5, is the presence of one or more nearly vertical lines
in the chaotic region, extending from very low to relatively high values of
${\chi}$. The lowest values of ${\chi}$ seem too small to be associated with
chaotic segments, but the upper values seem too large to be associated with
regular orbits. In fact, these lines contain two different classes of orbits,
namely regular orbits, for which ${\chi}(t)$ eventually decays
to zero, and confined chaotic orbits, trapped temporarily near a small
regular island, which eventually escape through one or more cantori to 
travel unimpeded through the stochastic sea. Viewed in configuration space,
the regular orbits corresponding to these lines correspond to periodic orbits
confined to an annulus or a ``figure-eight-shaped'' region. The confined 
chaotic orbits correspond to aperiodic orbits which, for a long time, are
trapped in almost the same region (whence follows the fact that they have
nearly the same ${\langle}{\tilde K}{\rangle}$ and ${\sigma}_{\tilde K}$ as
do the regular orbits), but eventually escape to probe the remaining chaotic
phase space regions.


Evidence for these assertions is provided in Fig.~6, which summarises an 
investigation of the longer time evolution of the initial conditions which
led to the near vertical line in Fig.~5b at 
${\sigma}_{\tilde K}{\;}{\approx}{\;}139$. 
A random sampling of $1261$ initial conditions with $E=10.0$ evolved for a 
total time ${\Delta}t=256$ led to $81$ segments with 
$138<{\sigma}_{\tilde K}<140$ and ${\chi}({\Delta}t)<0.1$. 
These $81$ initial conditions were subsequently integrated for a significantly
longer time, $t=8\times 256$, with ${\chi}(t)$ recorded at regular intervals,
and the resulting orbits partitioned into $8$ 
segments of length ${\Delta}t=256$, for which values ${\sigma}_{\tilde K}$
and ${\langle}{\tilde K}{\rangle}$ were computed. The recorded values of
${\chi}(t)$ were then analysed to extract short time Lyapunov exponents for
successive intervals of length ${\Delta}t=256$, the exponent for the 
segment extending from $t_{k}=k{\Delta}t$ to $t_{k}+{\Delta}t$ being defined
via the obvious relation (cf. [12])
$${\chi}({\Delta}t_{k})=
{(t_{k}+{\Delta}t){\chi}(t_{k}+{\Delta}t)-t_{k}{\chi}(t_{k})\over {\Delta}t}.
\eqno(18) $$

The tiny dots in all four panels of Fig. 6 exhibit the values of ${\chi}$
and ${\sigma}_{\tilde K}$ generated for the original ensemble of $1261$ orbits
evolved for a time $t=256$. The diamonds in Fig. 6a highlight the $81$ 
segments concentrated initially along the line near ${\sigma}_{\tilde K}=139$.
The diamonds in Fig.~6b show the values of ${\chi}$ and ${\sigma}_{\tilde K}$
derived from the same $81$ initial conditions for the period $256<t<512$. 
Figures 6c and d
extend the results to later intervals $512<t<768$ and $768<t<1024$. It is
clear that, as time elapses, the larger ${\chi}$ diamonds escape from the line
and move to other portions of the chaotic regions, whereas the smaller ${\chi}$
diamonds evolve closer to ${\chi}=0$.

The other complication common to the ${\Lambda}$ pattern is the existence of
smaller scale structures in the regular regions. As noted already, for short 
time exponents 
generated from a seed ${\delta}Z$ directed nearly in the $x$-direction, 
many/most regular segments fit along a curve with ${\chi}$ decreasing as one
moves away from the central chaotic region to much larger or smaller values
of ${\sigma}_{\tilde K}$. Alternatively, for more generic choices of 
${\delta}Z$, a plot of ${\chi}$ as a function of ${\sigma}_{\tilde K}$ 
exhibits an upper envelop that is is more nearly independent of 
${\sigma}_{\tilde K}$. However, in each case this upper curve is not the whole 
story. 

As illustrated in Figs. 5c and d, for generic values of ${\delta}z$ the
low ${\sigma}_{\tilde K}$ loop orbits exhibit two additional features, namely
(1) small scale oscillations in the ${\sigma}_{\tilde K}\,$-$\,{\chi}$ or 
${\langle}{\tilde K}{\rangle}\,$-$\,{\chi}$ plane, and (2) an excess 
probability of finding segments with especially low values of ${\chi}$ 
{\it closer to}, rather than further from, the central chaotic region. 
Alternatively, as illustrated in Figs. 5a and b, for ${\delta}z$ oriented more 
nearly in the $x$-direction one can identify various ``sub-families'' of 
regular orbits which,
for a fixed value of ${\sigma}_{\tilde K}$, tend to have much smaller values
of ${\chi}$ than other regular orbits with (nearly) the same value of
${\sigma}_{\tilde K}$. Thus, e.g., if one focuses on a fixed interval of 
values of ${\sigma}_{\tilde K}$ and compares regular orbits with larger values 
of ${\chi}$ that fall along the 
upper line with regular orbits with smaller values of ${\chi}$, he or she 
finds typically that there is a relatively clear cut difference between the 
two sets of orbits, as probed by the detailed shapes of their power spectra or,
in some cases, by the overall shape of the orbit. In many cases, the regular 
orbits that have smaller values of ${\chi}$ tend to be ``simpler'' or ``more 
regular'' in appearance. 
This is illustrated in Fig. 7, which exhibits two representative 
regular orbits extracted from the interval $51.0<{\sigma}_{\tilde K}<58.0$.
Figure 7a shows a typical low-${\chi}$ orbit, whereas Fig. 7b shows a typical
orbit lying along the upper curve. It is obvious that the higher-${\chi}$ 
orbit is (nearly) space filling within an annulus, whereas the lower orbit 
exhibits more structure.

The detailed patterns relating ${\chi}$ to ${\sigma}_{\tilde K}$ are different
for initial conditions evolved in the truncated Toda potential, but the data
still admit to a similar interpretation. At relative low energies, 
$E{\;}{\sim}{\;}2-8$, one sees patterns qualitatively similar to those observed
for $E>6$ or so in the dihedral potential. Alternatively, for somewhat
higher energies, $E>25$ or so, one sees instead a pattern reminiscent of
those observed at relatively low energies in the dihedral potential where,
in the absence of a significant population of large ${\sigma}_{\tilde K}$ box
orbits, one observes simply that, for chaotic segments, larger 
${\sigma}_{\tilde K}$ correlates with larger ${\chi}$. Superficially, the
behaviour for $E=20.0$, illustrated in Fig.~4c, might seem quite different 
from what is observed in the dihedral potential since here the
chaotic segments with the smallest values of ${\chi}$ are those furthest
from the regular regions. However, when placed in an appropriate context this 
behaviour is not surprising. At energies slightly above $E=20.0$, this low
${\chi}$ chaotic region merges into a relatively large regular island, which 
is well established by $E=25.0$. For $E=30.0$, one sees three large islands 
at low ${\sigma}_{\tilde K}$, the two corresponding to the lowest values of 
${\sigma}_{\tilde K}$ very similar qualitatively to the islands for $E=10.0$
in the dihedral potential associated with the two vertical lines.
\vskip .2in
\centerline{\bf C. Curvature and Complexity}
\vskip .1in
The preceeding shows that there is a strong correlation between the overall
stability or instability of an orbit segment, as probed by the value of a 
short time Lyapunov exponent ${\chi}(t)$, and the values of 
${\langle}{\tilde K}{\rangle}$ and ${\sigma}_{\tilde K}$ associated with that
segment. However, earlier work [12] has shown that, for chaotic
segments, there also exists a strong, nearly linear, correlation between 
${\chi}(t)$ and the {\it complexity} of the segment, as probed by the form of
its Fourier spectrum. One thus anticipates that there should exist correlations
between ${\langle}{\tilde K}{\rangle}$ or ${\sigma}_{\tilde K}$ and the
complexity of the segment.

As discussed more carefully in Ref. [12], $n(k)$, the complexity of an orbit 
segment at threshold $k$, can be defined in the following way: Given the 
values of the configuration space coordinates, say $x$ and $y$, at fixed 
time intervals ${\Delta}t$, sampled, e.g., ${\sim}{\;}5-10$ times per crossing
time, compute the power spectra, $|x({\omega})|$ and $|y({\omega})|$ in the
usual way [20]. Next determine the minimum number of frequencies, $n_{x}(k)$ 
and $n_{y}(k)$, required respectively to capture a fixed fraction $k$ of the
$x$- and $y$-powers. The total complexity is then defined as 
$$n(k)=n_{x}(k)+n_{y}(k). \eqno(19) $$

This notion of complexity is motivated by the idea that ``comparatively
regular'' segments have most of their power concentrated at or near a few 
special frequencies, whereas ``wildly chaotic'' segments have spectra with 
significantly
broader band power. Experience with ensembles of orbits evolved in several 
different potentials has indicated that values $k{\;}{\sim}{\;}0.9-0.95$ 
yield relative complexities in good agreement with subjective impressions
based on visual inspection of segments in the $x-y$ plane. The results 
described in this paper were derived from integrations where coordinates were
recorded at fixed intervals ${\Delta}t=0.125$ for a total time $t=256$, this
leading to $2048$ points and hence a Fourier series with $2048$ frequencies.

As stated already, for orbit segments in both the dihedral and truncated Toda 
potentials, as well as for other systems [12], 
there is a strong, nearly linear correlation between ${\chi}(t)$ and quantities
like $n(0.9)$ or $n(0.95)$. For ensembles where ${\chi}$ and $n(k)$ both
assume a broad range of values, the rank correlation typically assumes a
value ${\cal R}({\chi},n(k))>0.85-0.9$. If, alternatively, most of the 
segments 
are concentrated in the same part of the ${\chi}-n(k)$ plane, ${\cal R}$ can
be smaller but still remains appreciable. 

This correlation can be understood intuitively in terms of the fact that 
chaotic orbit segments often appear visually to be comprised of ``pieces'' of 
various regular orbits that exist at the same energy [16]. Consider, e.g., 
orbits 
with $E=10.0$ in the dihedral potential. Here one finds that chaotic segments 
with values of ${\sigma}_{\tilde K}$ near those appropriate for the two large
regular regions tend overall to look more regular, and to have smaller
complexities, than segments located at values of ${\sigma}_{\tilde K}$ further
from the regular regions. Moreover, chaotic segments near the low 
${\sigma}_{\tilde K}$ regular region tend to look ``loopy'' and to have power 
spectra dominated by features appropriate for the loop orbits that exist at 
small ${\sigma}_{\tilde K}$; and similarly, chaotic segments near the high 
${\sigma}_{\tilde K}$ regular region tend to look ``boxy'' and to have spectra 
similar to the high ${\sigma}_{\tilde K}$ box orbits. 

The degree to which a quantity like $n(0.9)$ correlates with ${\chi}$, both
in itself and in relation to other quantities like 
${\langle}{\tilde K}{\rangle}$ or ${\sigma}_{\tilde K}$, can be gauged from 
Fig.~8, which exhibits data generated in the dihedral potential for $E=10.0$
and $E=1.0$. The two left panels exhibit plots of ${\chi}$ as functions of
${\langle}{\tilde K}{\rangle}$, whereas the two right panels plot $n(0.9)$
as a function ${\langle}{\tilde K}{\rangle}$. It is clear that, at least in
the chaotic regions, the plots of ${\chi}$ and $n(0.9)$ exhibit the same
basic features, although the plots of $n(0.9)$ seem more ``blurred.''

There is, however, a significant difference for the regular orbits. Plots
of $n(0.9)$ (or $n(0.95)$) as a function of ${\sigma}_{\tilde K}$ or
${\langle}{\tilde K}{\rangle}$ exhibit large systematic oscillations not 
manifested 
in the plots of ${\chi}$, whereas the small scale structures present in the
plots of ${\chi}$ (cf. Fig. 5) are absent in plots of $n(0.9)$. Consider, e.g.,
$E=10.0$. Here all the segments at large ${\langle}{\tilde K}{\rangle}$ are 
regular loops, but the loops near ${\langle}{\tilde K}{\rangle}=175$ are 
clearly special in that $n(0.9)$ assumes values that are particularly small.
This reflects the fact that, in this region, peaks in the spectra 
$|x({\omega})|$ and $|y({\omega})|$ are much sharper than at somewhat larger 
and smaller values of ${\langle}{\tilde K}{\rangle}$. 

This is a finite sampling effect, reflecting the fact that the orbits segments
were integrated for a relatively short time $t=256$. If the total time $t$
is changed, but data are still recorded and analysed at $2048$ equally spaced
intervals, one finds that, as would be expected, the values of
${\langle}{\tilde K}{\rangle}$ and ${\sigma}_{\tilde K}$ computed for regular
segments remain essentially
unchanged. By contrast, however, the details of the oscillations can be
alterred significantly. Thus, e.g., for $E=1.0$ the location of the dip near
${\langle}{\tilde K}{\rangle}=17.5$ will move and other minima can appear.
Moreover, one observes that, as $t$ increases, the computed complexities
decrease and the amplitudes of the oscillations damp.

One final correlation remains to be stated. To say that a regular orbit 
is close to the chaotic region is a statement
{\it both} about its location (say) in the ${\sigma}_{\tilde K}\,$-$\,{\chi}$ 
plane {\it and} its actual location in configuration space, as probed, e.g., 
by a surface of section. For example, orbits in the large regular regions with 
values of ${\langle}{\tilde K}{\rangle}$ and ${\sigma}_{\tilde K}$ further from
the central chaotic region tend systematically to lie closer to the center of
a regular island than do regular segments with values of 
${\langle}{\tilde K}{\rangle}$ and ${\sigma}_{\tilde K}$ closer to the central
chaotic region. Similarly, the vertical lines associated with regular and
confined chaotic orbits are associated with smaller regular islands 
embedded in the stochastic sea.

These facts are illustrated in Fig. 9. Here the small dots represent 
successive sections generated from a single ``unconfined'' chaotic initial 
condition that systematically avoids the phase space regions near the two 
large regular islands. The remaining symbols represent different initial 
conditions corresponding to regular and ``confined'' chaotic orbits generated
from a uniform sampling of the $y\,$-$\,p_{y}$ plane. The smaller of the
two large islands, located near $y=p_{y}=0$ and populated with stars, 
corresponds to box orbits. The larger, displaced from $y=0$, corresponds 
instead to loop orbits. Here the initial conditions were binned into three
classes. The diamonds represent segments with ${\sigma}_{\tilde K}$ assuming
particularly small values, whereas the squares represent values of 
${\sigma}_{\tilde K}$ especially close to the chaotic region. The triangles
correspond to intermediate values. The remaining pluses and crosses 
represent initial conditions which led to the two nearly vertical lines
exhibited in Fig. 5. 
\vskip .2in
\centerline{\bf III. SUMMARY AND INTERPRETATION}
\vskip .1in
Several general trends emerge from the experiments described in the preceeding
Section. Most striking and fundamental, perhaps, is the fact that, for an 
ensemble of orbit segments of fixed energy $E$, a plot of the dispersion 
${\sigma}_{\tilde K}$ as a function of the mean ${\langle}{\tilde K}{\rangle}$ 
usually assumes a very simple form. In general, regular and chaotic segments 
appear to coexist along a single curve in the 
${\langle}{\tilde K}{\rangle}\,$-$\,{\sigma}_{\tilde K}$ plane, although for 
both potentials one sees more complicated structures at very low energies. 
The fact that chaotic segments fit along a single curve indicates that, even
though they are aperiodic, they still exhibit significant statistical 
regularities.

Plots of the short time Lyapunov exponent ${\chi}(t)$ as a function of 
${\langle}{\tilde K}{\rangle}$ or ${\sigma}_{\tilde K}$ also exhibit simple,
distinctive patterns. For example, one finds that, largely independent of the 
initial
perturbation ${\delta}Z$ used in calculating ${\chi}$, chaotic segments with
values of ${\langle}{\tilde K}{\rangle}$ (or ${\sigma}_{\tilde K}$) far from
the regions associated with regular orbits tend to have larger values of
${\chi}(t)$ than do chaotic segments with values of 
${\langle}{\tilde K}{\rangle}$ (or ${\sigma}_{\tilde K}$) nearer the regular
regions. For regular orbits, the computed values of ${\chi}(t)$, and hence
any correlations with ${\langle}{\tilde K}{\rangle}$ or ${\sigma}_{\tilde K}$,
depend more sensitively on the initial perturbation. For most initial 
perturbations, it seems that, overall, the typical ${\chi}(t)$ is nearly 
independent of ${\langle}{\tilde K}{\rangle}$ and ${\sigma}_{\tilde K}$. 
However, for some classes of perturbations one finds instead that segments 
with values of ${\langle}{\tilde K}{\rangle}$ (or ${\sigma}_{\tilde K}$) 
further from the chaotic regions tend to have smaller values of ${\chi}$ than 
do regular segments with values closer to the chaotic region. 

The correlations between ${\chi}(t)$ and ${\langle}{\tilde K}{\rangle}$ (or 
${\sigma}_{\tilde K}$) observed for chaotic orbits are very similar to the 
observed correlations between the complexity $n(k)$ and 
${\langle}{\tilde K}{\rangle}$ (or ${\sigma}_{\tilde K}$). This reflects the 
fact that chaotic segments with values of ${\langle}{\tilde K}{\rangle}$ and 
${\sigma}_{\tilde K}$ closer to the regular regions tend to look more regular 
and, consequently, to have less complex Fourier spectra than do orbits with
values further from the regular region.

One other common feature, observed for both complexities and short time 
exponents, is the presence of one or more nearly vertical lines in (say) the 
${\sigma}_{\tilde K}\;$-$\,{\chi}$ or ${\sigma}_{\tilde K}\;$-$\,n(k)$ plane. 
These are comprised of two distinct types of segments, namely segments of
regular orbits trapped forever in an island of stability by invariant 
{\it KAM} tori and ``confined'' chaotic orbits, trapped temporarily by cantori 
in a similar configuration space region which, however, eventually diffuse 
through the cantori to probe the remaining portions of the chaotic sea. 

Many of these qualitative features can be reproduced in a much simpler context,
where the linearised scalar Jacobi equation (11) is replaced by a linear 
Matthieu equation of the form [21]
$${d^{2}{\xi}_{1}\over dt^{2}}
=-({\omega}^{2}+{\gamma}\sin\,t){\xi}_{1} , \eqno(20)$$
where ${\omega}^{2}$ and ${\gamma}$ play the roles, respectively, of 
${\langle}{\tilde K}{\rangle}$ and ${\sigma}_{\tilde K}$.

If one studies solutions to eq. (20) as a function of ${\omega}^{2}$ and 
${\gamma}$
(or, by analogy, ${\langle}{\tilde K}{\rangle}$ and ${\sigma}_{\tilde K}$)
he or she finds that the ${\omega}\,$-$\,{\gamma}$ plane is partitioned into a
large number of disjoint regions, corresponding respectively to ``regular'' and
``chaotic'' orbits. In the regular regions, it is apparent that, in the 
$t\to\infty$ limit, the analogue of the ordinary short time Lyapunov exponent,
$${\chi}(t)={1\over t}{\;}
\ln {\Biggl[}{||{\delta}Z(t)||\over ||{\delta}Z(0)||}{\Biggr]}, \eqno(21) $$
with $||{\delta}Z||^{2}=|{\xi}_{1}|^{2}+|{\dot\xi}_{1}|^{2}$, eventually
converges towards zero. Alternatively, in the chaotic regions ${\chi}(t)$
converges towards a time-independent positive value. It follows that a generic
curve passing through the ${\omega}\,$-$\,{\gamma}$ (or
${\langle}{\tilde K}{\rangle}\,$-$\,{\sigma}_{\tilde K}$) plane, analogous to 
the curves exhibited in Figs. 1 and 2, will include both regular and 
chaotic regions with sharp transitions between the two.

For values of ${\omega}$ and ${\gamma}$ far from the chaotic portions of the
curve, one finds typically that, for reasonably large $t$, the computed value
of ${\chi}$ is relatively insensitive to the detailed perturbation and to the
exact values of ${\omega}$ and ${\gamma}$. However, for values of ${\omega}$
and ${\gamma}$ closer to the chaotic regions, more complicated patterns can
arise. Viewed in an asymptotic $t\to\infty$ limit, the transition from regular 
to chaotic is abrupt. However, the transition is smooth in the sense that, when
evaluated for some finite time $t$, the short time ${\chi}(t)$ computed for
``chaotic'' values of ${\omega}$ and ${\gamma}$ close to the regular regions
is typically smaller than the ${\chi}(t)$ computed for values that are further
from the regular region.
This is consistent
with the fact, manifested in Figs. 3 - 5, that orbits with values of
${\langle}{\tilde K}{\rangle}$ and ${\sigma}_{\tilde K}$ further from the
regular regions tend to have larger short time Lyapunov exponents. Given this 
observation, it is also easy
to understand (cf. Fig. 6) why, for ``sticky'' chaotic segments confined near 
regular regions by cantori, the computed ${\chi}(t)$ tends to be smaller than
for other chaotic segments that travel further from the regular regions.

A detailed discussion of the sense in which the onset of chaos can be 
understood in terms of a more general Hill equation has been presented by 
Cerruti-Sola and Pettini [7] in the context of their analysis of individual 
orbits.

The work described in this paper has established  the existence of a strong
correlation between such geometric properties of chaotic orbit segments
as ${\langle}{\tilde K}{\rangle}$ and ${\sigma}_{\tilde K}$ and the
short time Lyapunov exponent ${\chi}$. However, one might expect that
this local, short time analyses could also be extended to establish 
a global connection between moments of ${\tilde K}$, as defined along an
infinite chaotic geodesic, and the ordinary Lyapunov exponent 
${\chi}_{\infty}$, as defined
in a $t\to\infty$ limit. Casetti {\it et al} [22] have shown that, at least
for systems with a large number $N$ of degrees of freedom, such a connection
does indeed exist. Specifically, by assuming that the curvature experienced
by a chaotic orbit can be approximated as a random process, characterised
by a mean ${\langle}K{\rangle}$ and a dispersion ${\sigma}_{K}$, they 
obtained an analytic approximation to the largest Lyapunov exponent which, in
at least some cases, agrees extremely well with numerical computations.

It would seem unlikely that such a simple Gaussian approximation, which 
these authors motivate in the large $N$ limit, will work well for the very
special case $N=2$. However, one might nevertheless hope that, in some
fashion, the value of the positive Lyapunov exponent for a two-dimensional
system can again be related to the moments of $K$ or ${\tilde K}$
evaluated along the chaotic orbit. This possibility is currently
under investigation.
\vskip .5in
\centerline{\bf ACKNOWLEDGMENTS}
\vskip .1in
I am grateful to Haywood Smith and, especially, Marco Pettini, for
helpful comments on preliminary versions of this paper. This research
was supported in part by the National Science 
Foundation Grant PHY92-03333. Some of the preliminary analysis was performed
by Barbara L. Eckstein, who has been supported by NASA through the Florida 
Space Grant Consortium. Some of the computations were facilitated by computer 
time made available through the
Research Computing Initiative at the Northeast Regional Data Center (Florida)
by arrangement with IBM.
\vfill\eject
\def\baselinestretch{1.2}   
\par\noindent
[1] Cf. L. D. Landau and E. M. Lifshitz, {\it Mechanics} (Pergamon, Oxford, 
1960) or V. I. Arnold, {\it Mathematical Methods of Classical Mechanics} 
(Springer, Berlin, 1978).
\par\noindent
[2] Cf. E. Hopf, Trans. Am. Math. Soc. {\bf 39}, 229 (1936).
\par\noindent
[3] D. V. Anosov, Trudy Mat. Inst. Steklov {\bf 90}, 1 (1967).
\par\noindent
[4] V. G. Gurzadyan and G. K. Savvidy, Astron. Astrophys. {\bf 160}, 203 
(1986).
\par\noindent
[5] H. E. Kandrup, Physica {\bf A 169}, 73 (1990); Astrophys. J. {\bf 364}, 420
(1990).
\par\noindent
[6] M. Pettini, Phys. Rev. {\bf E 47}, 828 (1993).
\par\noindent
[7] M. Cerruti-Sola and M. Pettini, Phys. Rev. {\bf E 53}, 179 (1995).
\par\noindent
[8] M. H\'enon and C. Heiles, Astron. J. {\bf 69}, 73 (1964).
\par\noindent
[9] Cf. P. Grassberger, R. Badii, and A. Politi, J. Stat. Phys. {\bf 51}, 135
(1988).
\par\noindent
[10] Cf. J. N. Mather, Topology {\bf 21}, 457 (1982); R. S. MacKay, J. D. 
Meiss, and I. C. Percival, Phys. Rev. Lett. {\bf 52}, 697 (1984).
\par\noindent
[11] M. E. Mahon, R. A. Abernathy, B. O. Bradley, and H. E. Kandrup, Mon. Not.
R. Astr. Soc. {\bf 275}, 443 (1995).
\par\noindent
[12] H. E. Kandrup, B. L. Eckstein, and B. O. Bradley, Astron. Astrophys.
{\bf 320}, 65 (1997).
\par\noindent
[13] D. Armbruster, J. Guckenheimer, and S. Kim, Phys. Lett. {\bf A 140}, 416
(1989).
\par\noindent
[14] M. Toda, J. Phys. Soc. Japan {\bf 22}, 431 (1967).
\par\noindent
[15] This is of especial interest physically given the fact that, for a system 
like a galaxy, $100t_{cr}$ corresponds to a period of time comparable to the 
age of the Universe.
\par\noindent
[16] In certain cases, e.g., for chaotic orbits trapped
by cantori near regular islands, it is not easy to determine whether or not a 
given segment is in fact chaotic. To make a precise determination in such 
potentially ambiguous cases, the segment was integrated for significantly 
longer times $t=2048$ to see whether the computed short time Lyapunov exponent 
${\chi}(t)$ eventually begins to increase and/or whether, in configuration
space, the segment eventually moves away from the regular island. When even 
such longer time integrations were inconclusive, the orbits were reintegrated
in the presence of very weak additive white noise since (cf. S. Habib, H. E. 
Kandrup, and M. E. Mahon, Phys. Rev. {\bf E 53}, 5473 [1996]; Astrophys. J.
{\bf 480}, 155 [1997]) such perturbations
tend to decrease dramatically the time scale on which chaotic orbits diffuse
through cantori without allowing regular orbits to breach true {\it KAM} tori.
\par\noindent
[17] H. E. Kandrup and B. L. Eckstein, Ann. N. Y. Acad. Sci. {\bf 808}, 139 
(1997).
\par\noindent
[18] Cf. G. Bennetin, L. Galgani, and J.-M. Strelcyn, Phys. Rev. {\bf A 14},
2338 (1976).
\par\noindent
[19] Cf. G. Contopoulos, Astron. J. {\bf 76}, 147 (1971) or R. S. Shirts and
W. P. Reinhart, J. Chem. Phys. {\bf 77}, 5204 (1982).
\par\noindent
[20] Cf. W. H. Press, B. P. Flannery, S. A. Teukolsky, and W. T. Vetterling,
{\it Numerical Recipes}, 2nd edition (Cambridge University Press, Cambridge,
1992). 
\par\noindent
[21] Cf. E. T. Whittaker and G. N. Watson, {\it A Course of Modern Analysis}
(Cambridge University Press, Cambridge, 1965).
\par\noindent
[22] L. Casetti, C. Clementi, and M. Pettini, Phys. Rev. {\bf E 54}, 5969 
(1996).
\vfill\eject
\centerline{\bf FIGURE CAPTIONS}
\vskip .2in
\par\noindent
FIG. 1 -- The mean ${\langle}{\tilde K}{\rangle}$ and dispersion 
${\sigma}_{\tilde K}$ for orbit ensembles with eight different energies $E$ 
evolved in the dihedral potential for a total time $t=256$. (a) $E=-0.05$. 
(b) $E=0.05$. (c) $E=0.25$. (d) $E=1.0$. (e) $E=2.0$. (f) $E=4.0$. (g)
$E=8.0$. (h) $E=20.0$. 
\vskip .1in
\par\noindent
FIG. 2 -- The mean ${\langle}{\tilde K}{\rangle}$ and dispersion 
${\sigma}_{\tilde K}$ for orbit ensembles with four different energies $E$ 
evolved in the truncated Toda potential for a total time $t=256$. 
(a) $E=0.5$. (b) $E=3.0$. (c) $E=20.0$. (d) $E=30.0$ 
\vskip .1in
\par\noindent
FIG. 3 -- The short time Lyapunov exponent ${\chi}$ plotted as a function
of ${\sigma}_{\tilde K}$ for the same orbit 
ensembles as in Fig. 1. 
\vskip .1in
\par\noindent
FIG. 4 -- The short time Lyapunov exponent ${\chi}$ plotted as a function
of ${\sigma}_{\tilde K}$ for the same orbit 
ensembles as in Fig. 2. 
\vskip .1in
\par\noindent
FIG. 5 -- (a) Short time Lyapunov exponents ${\chi}(t)$, generated with an 
initial perturbation ${\delta}x=10^{-10}$ and plotted as functions of 
${\sigma}_{\tilde K}$ for an ensemble of orbit segments with $E=10.0$ evolved 
in the dihedral potential for a total time $t=256$. Only values of 
${\chi}<0.16$ are shown. (b) The same ${\chi}(t)$'s plotted as a function of
${\langle}{\tilde K}{\rangle}$. (c) The analogue of FIG. 5a, generated for
the same ensemble, but now allowing for an initial perturbation 
${\delta}y=10^{-10}$. 
(d) The analogue of FIG. 5b, generated for an initial perturbation 
${\delta}y=10^{-10}$.
\vskip .1in
\par\noindent
FIG. 6 -- The evolution of regular and confined chaotic orbits for $E=10.0$
in the dihedral potential, with ${\chi}$ plotted as a function of 
${\sigma}_{\tilde K}$. The dots represent analogues of FIG. 3, with a
total integration time $t=256$. The diamonds represent the same initial 
conditions analysed over different time intervals. (a) $0<t<256$. (b)
$256<t<512$. (c) $512<t<768$. (d) $768<t<1024$.
\vskip .1in
\par\noindent
FIG. 7 -- Two orbits with $E=10.0$ evolved in the dihedral potential for a
time $t=256$. (a) An orbit with ${\langle}{\tilde K}{\rangle}=183.4$, 
${\sigma}_{K}=56.4$, $Q=0.308$, and ${\chi}=0.0051$. (b) An orbit with 
${\langle}{\tilde K}{\rangle}=181.6$, ${\sigma}_{K}=62.7$, $Q=0.345$, and 
${\chi}=0.0116$. 
\vskip .1in
\par\noindent
FIG. 8 -- (a) The short time exponent ${\chi}$ plotted as a function of 
${\langle}{\tilde K}{\rangle}$ for a collection of orbits with $E=10.0$ 
evolved in the dihedral potential for a total time $t=256$. (b) The complexity
$n(0.9)$ plotted as a function of ${\langle}{\tilde K}{\rangle}$ for the
same set of segments. (c) The analogue of FIG. 8 (a) for a collection of
segments with $E=1.0$. (d) The analogue of FIG. 8 (b) for the segments of
FIG. 8 (c).
\vskip .1in
\par\noindent
FIG. 9 -- An $x$-$p_{x}$ surface of section for $E=10.0$ in the dihedral 
potential. The dots derive from a single unconfined chaotic orbit evolved 
for $2000$ intersections. The remaining points represent some of the initial 
conditions used to generate FIGS. 5 and 6. Diamonds represent loop orbits 
with ${\sigma}_{\tilde K}<72.0$. Triangles represent loop orbits with 
$72.0<{\sigma}_{\tilde K}<90.0$. Squares represent loop orbits with 
$90.0<{\sigma}_{\tilde K}<110.0$. Stars represent box orbits with 
${\sigma}_{\tilde K}>141.0$. Pluses represent points along the near-vertical
line with $120.0<{\sigma}_{\tilde K}<122$ and ${\chi}<0.08$. Crosses represent 
points along the line with $138.0<{\sigma}_{\tilde K}<140.0$ and ${\chi}<0.1$. 
\vfill\eject
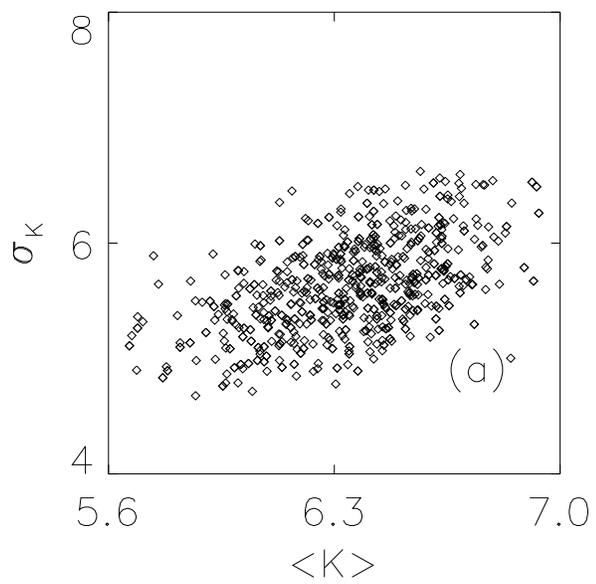
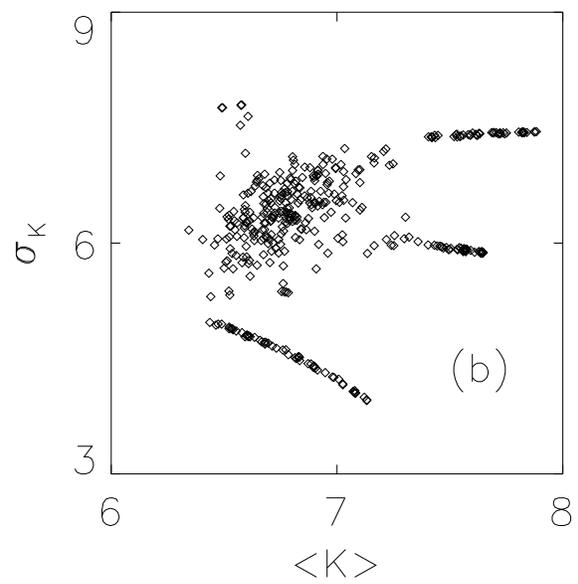
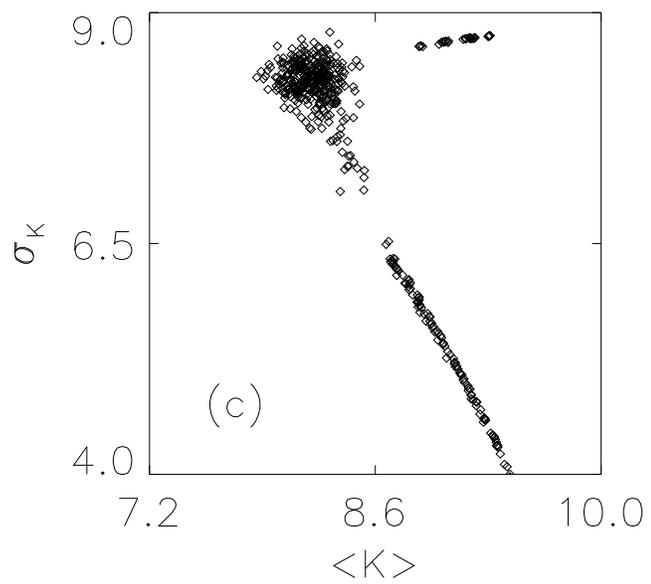
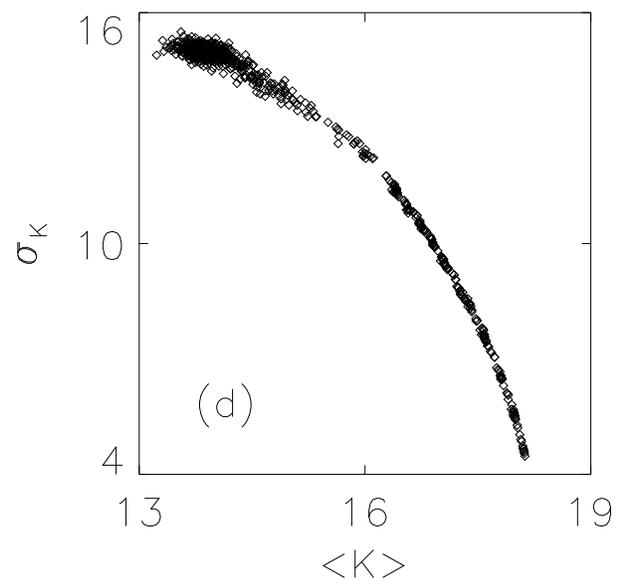
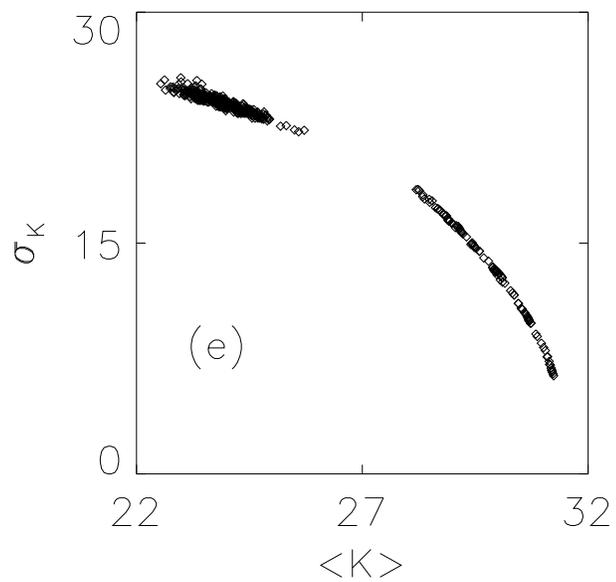
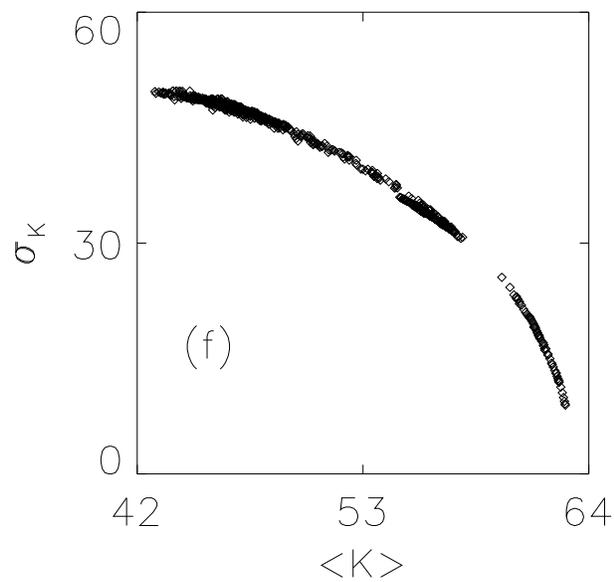
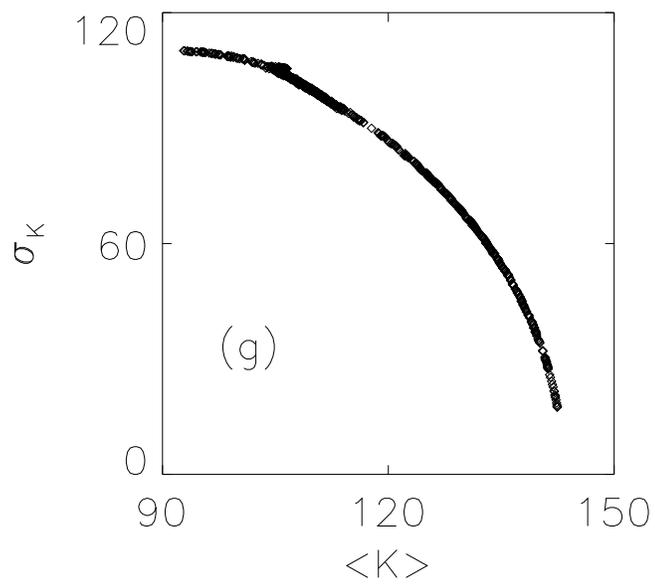
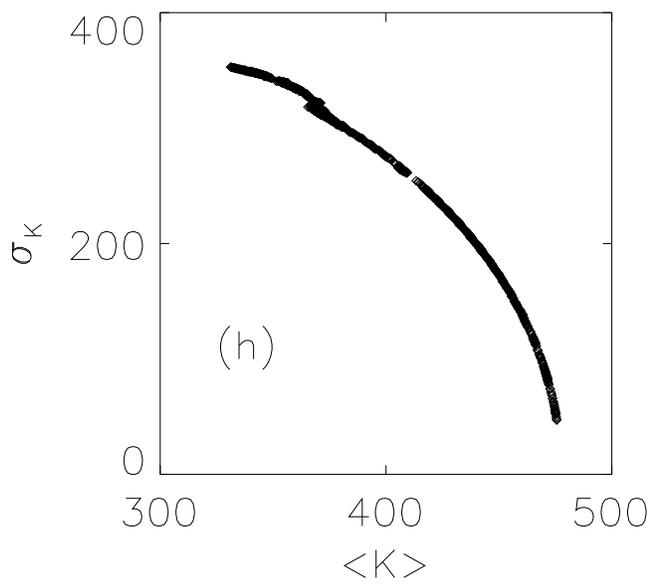
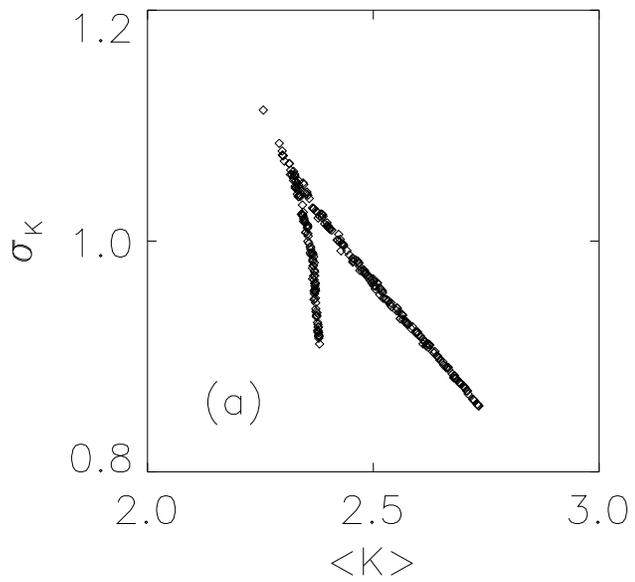
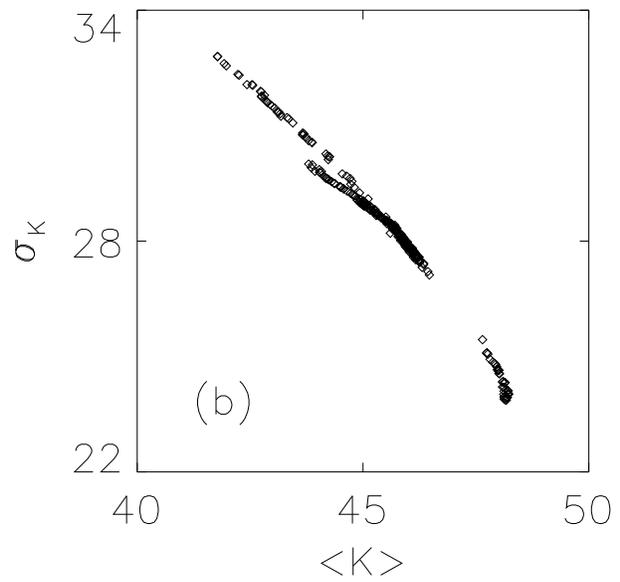
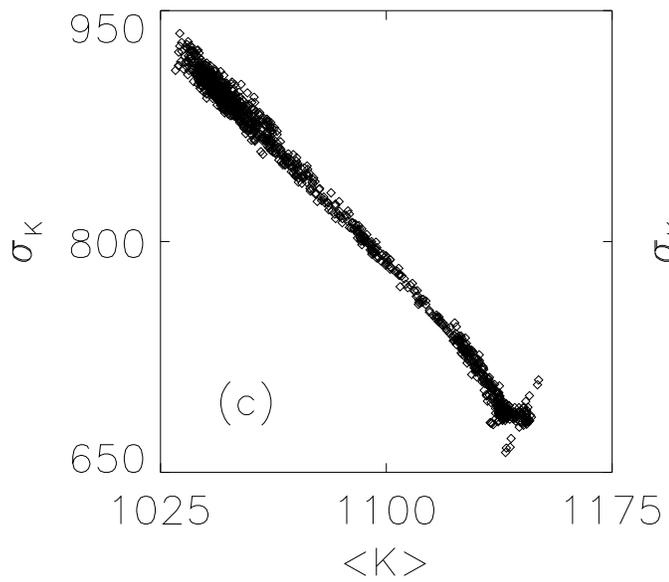
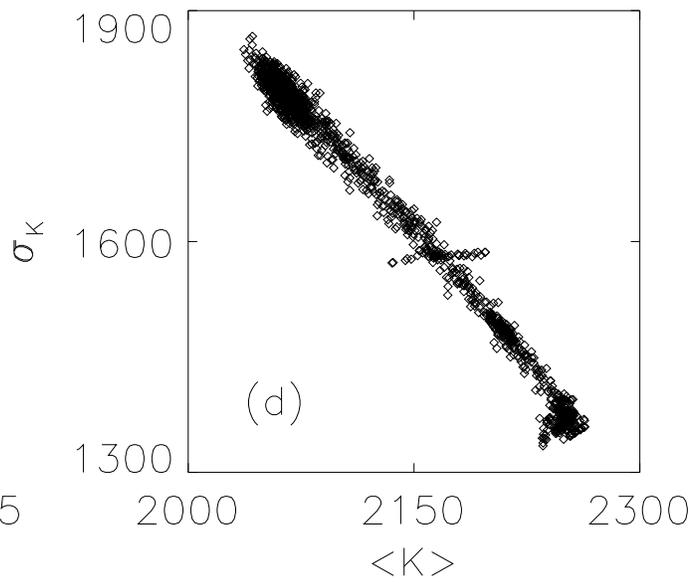
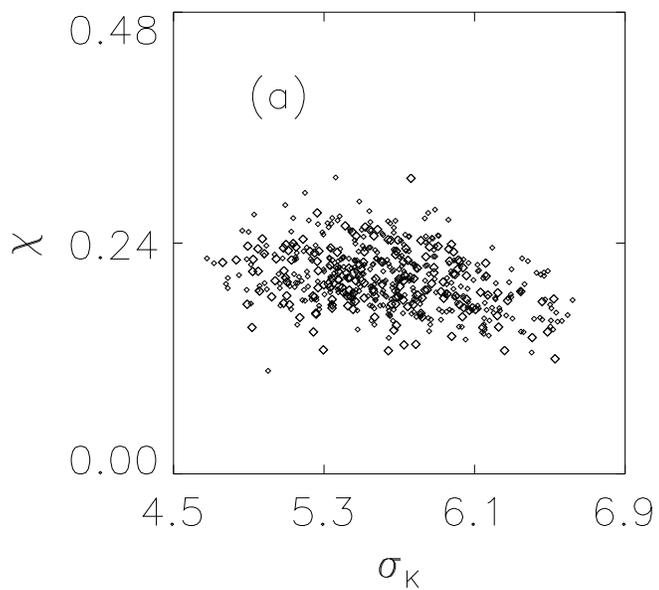
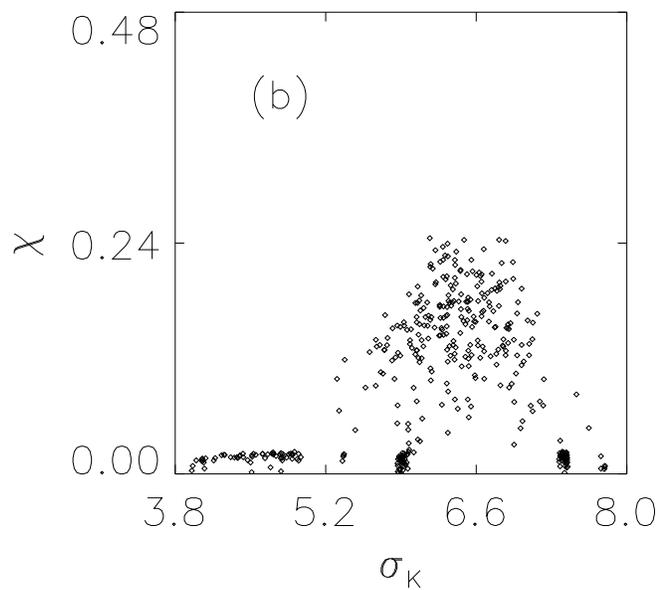
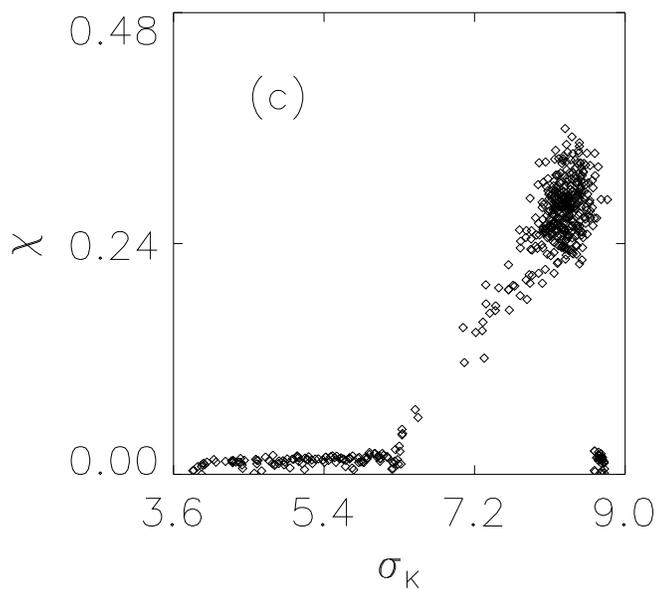
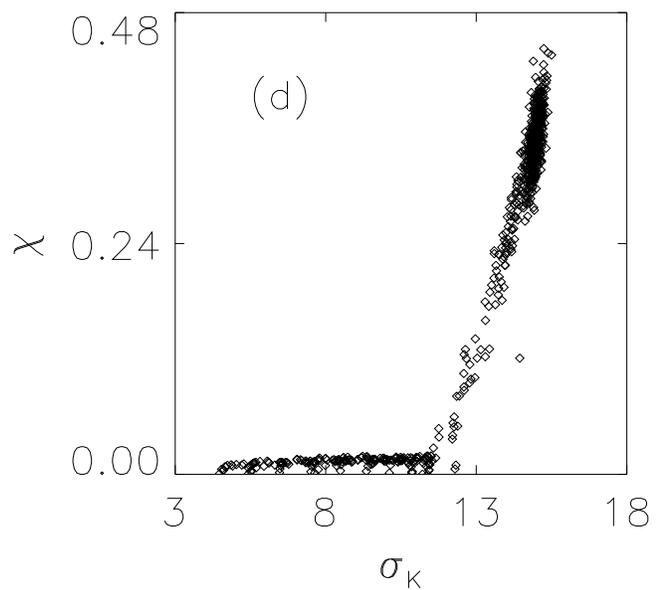
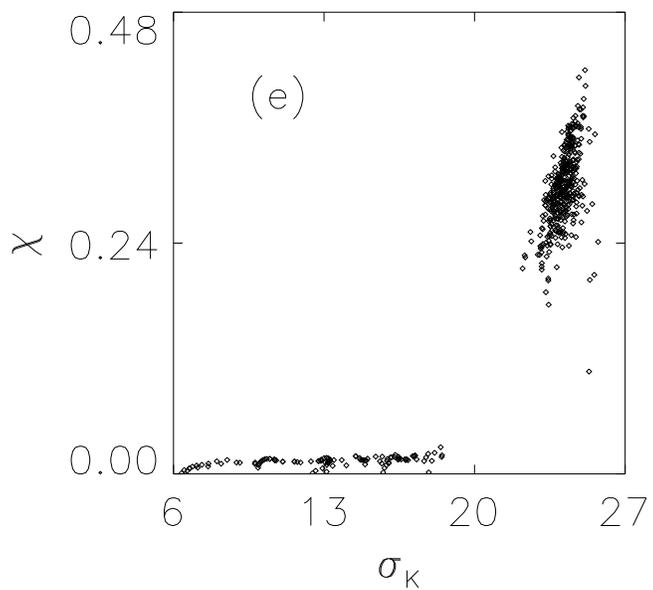
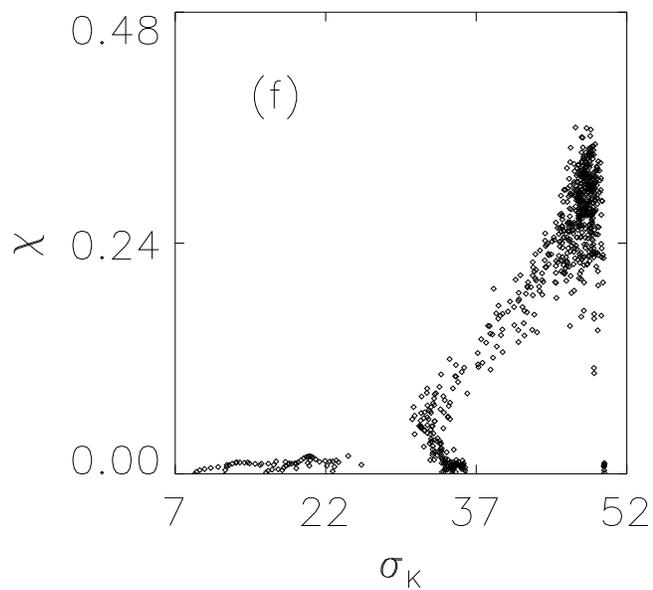
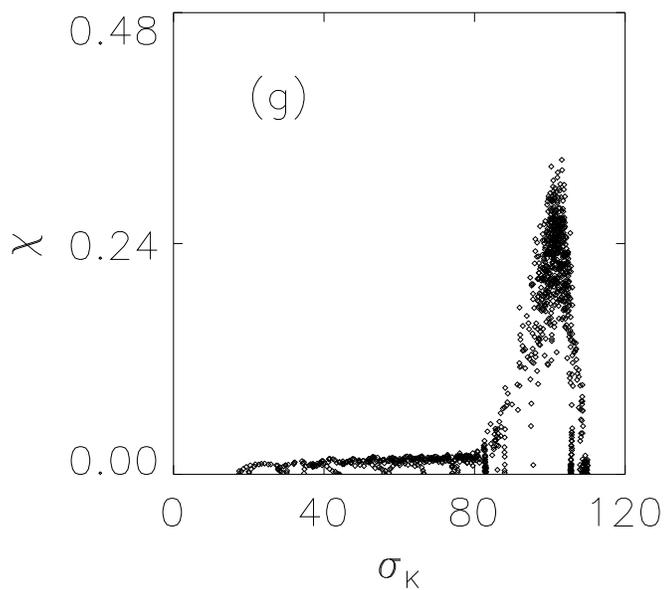
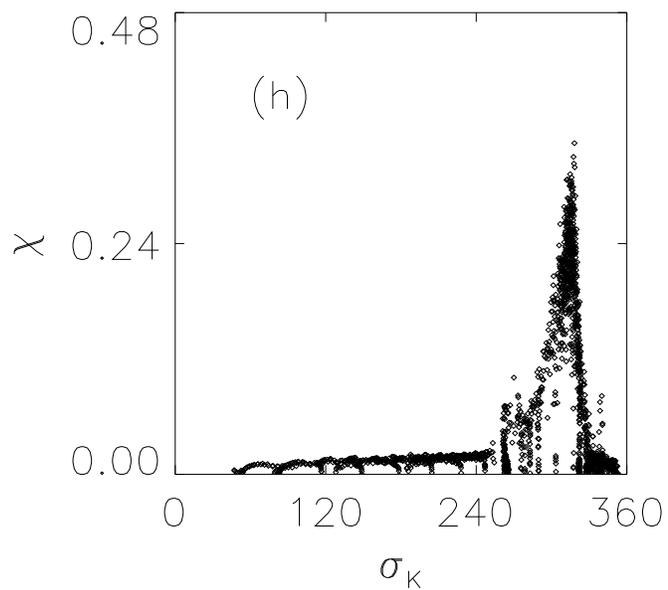
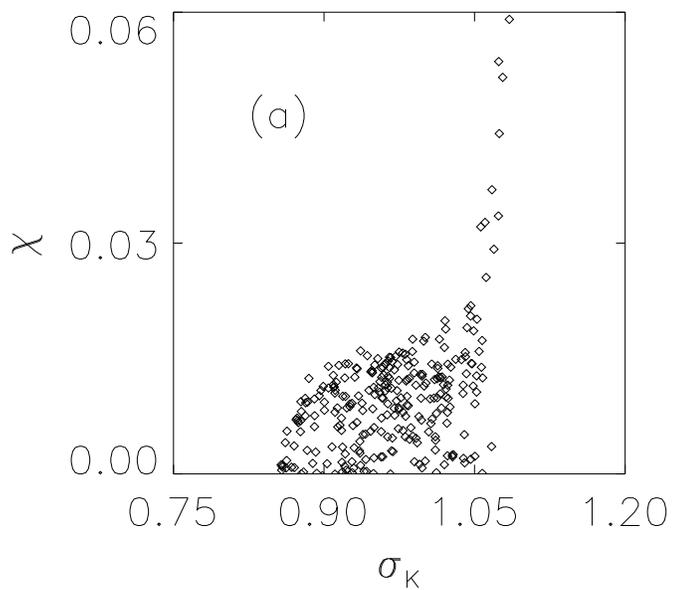
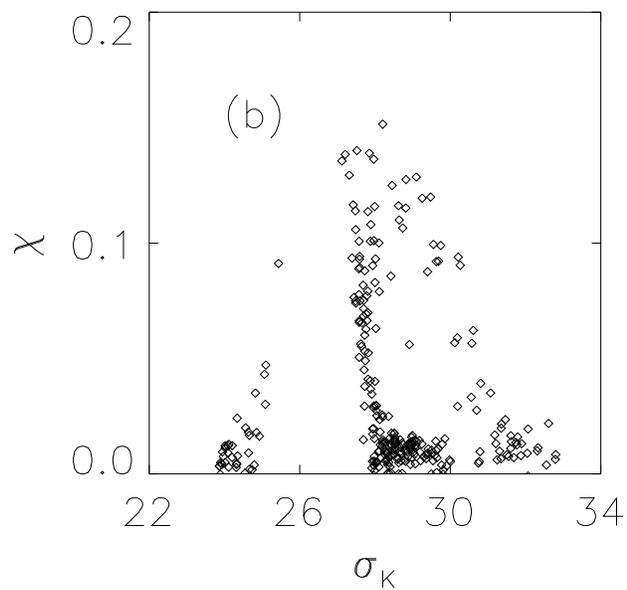
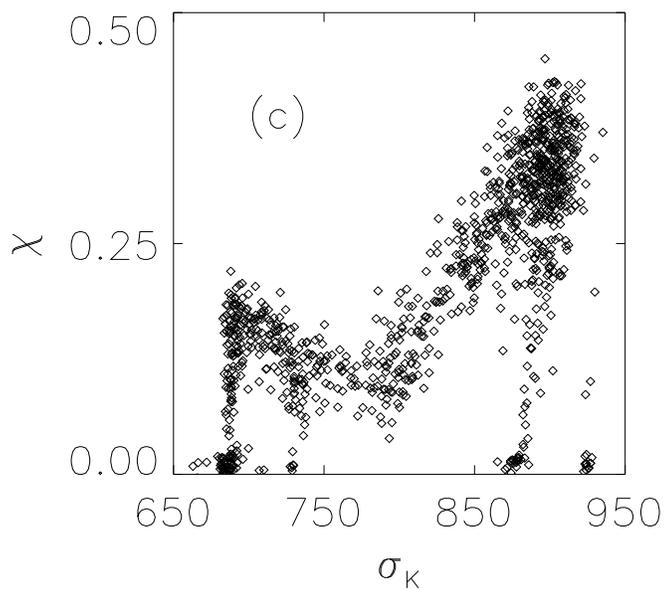
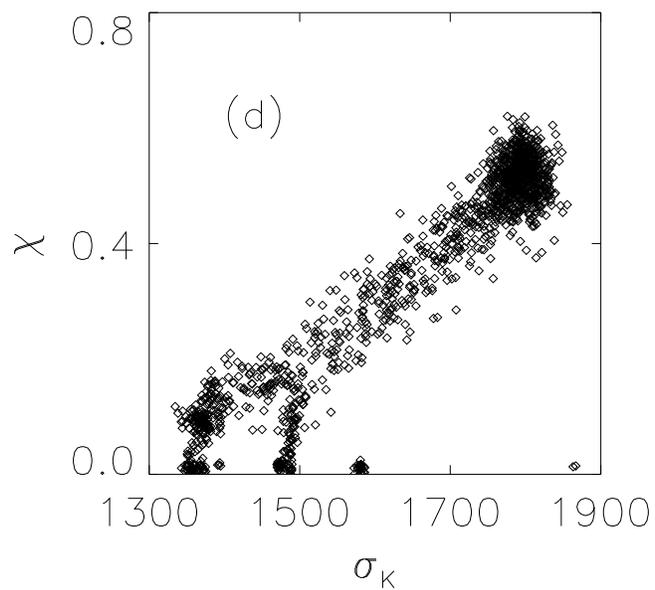
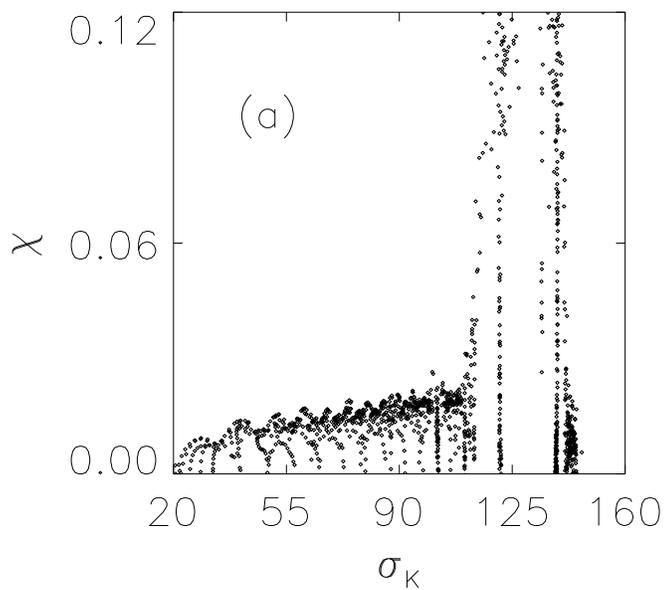
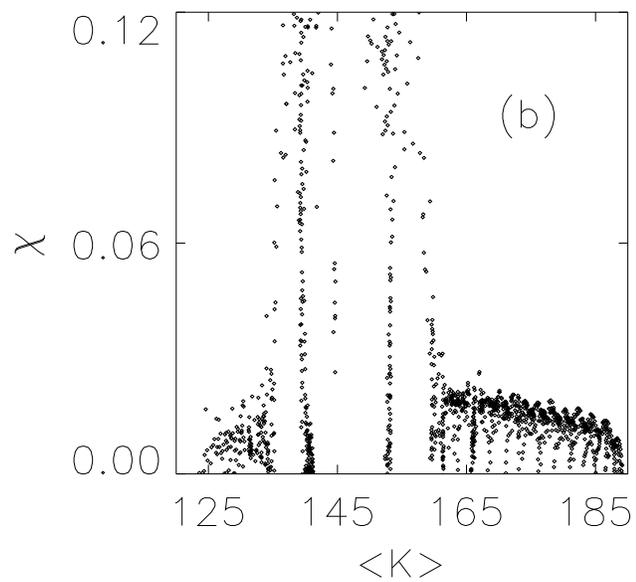
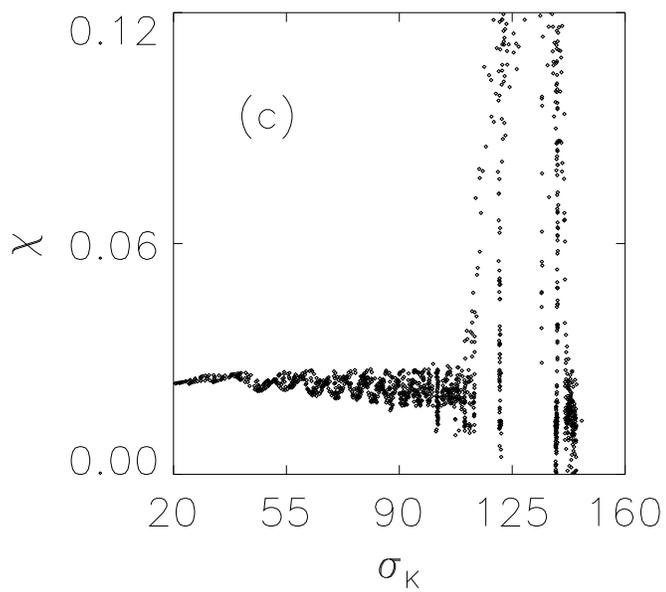
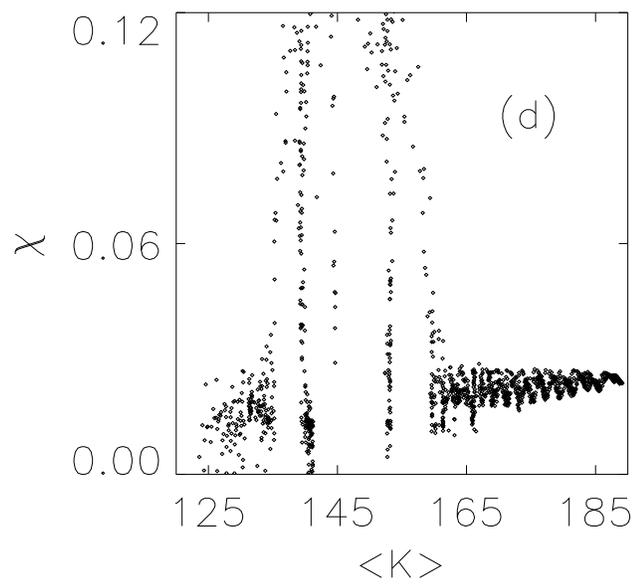
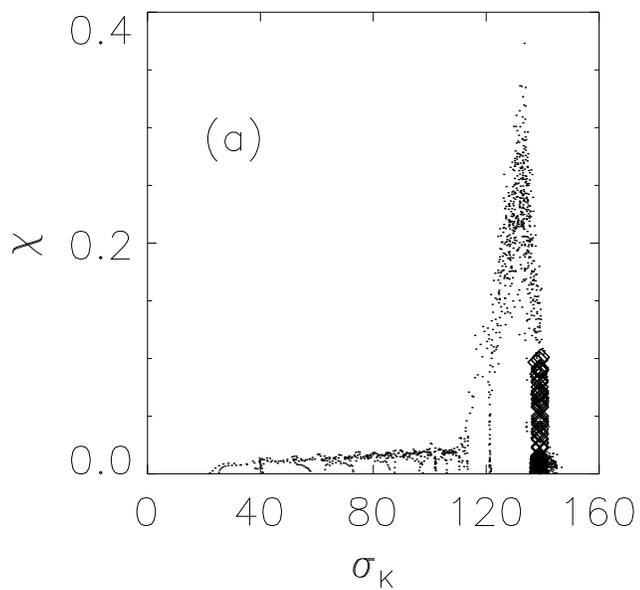
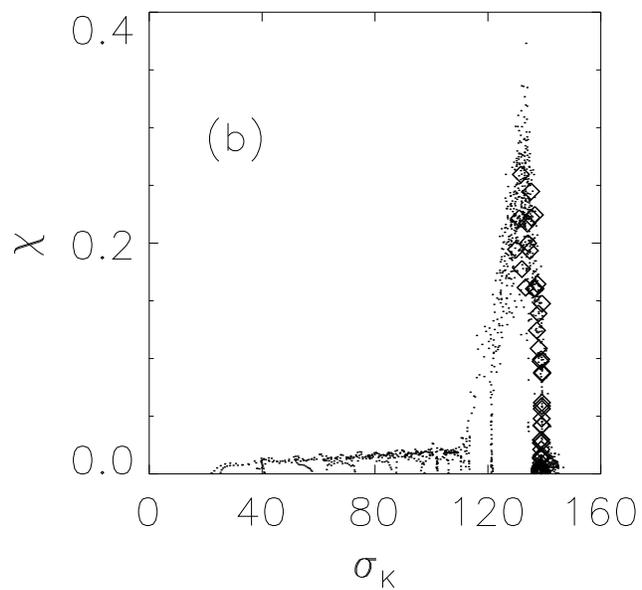
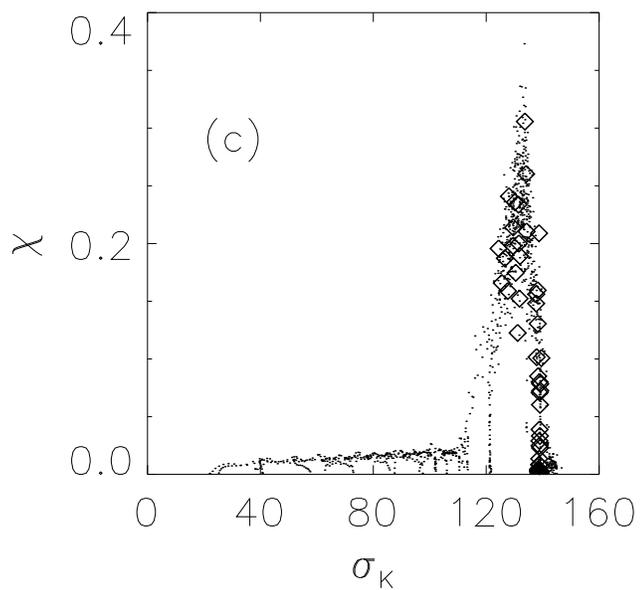
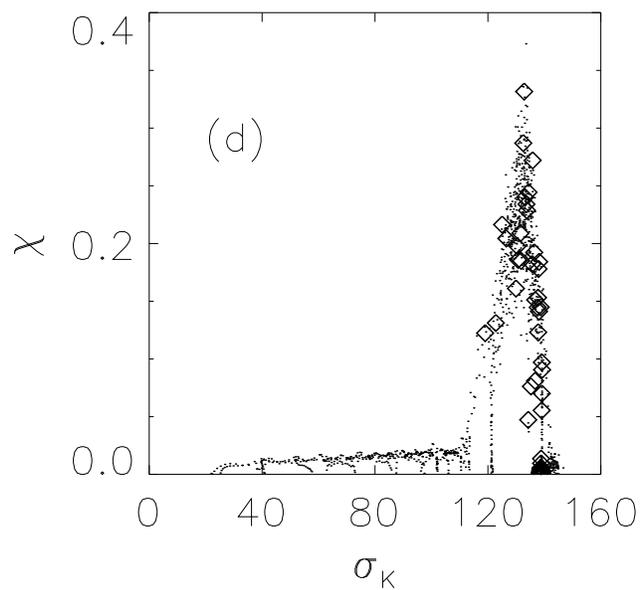
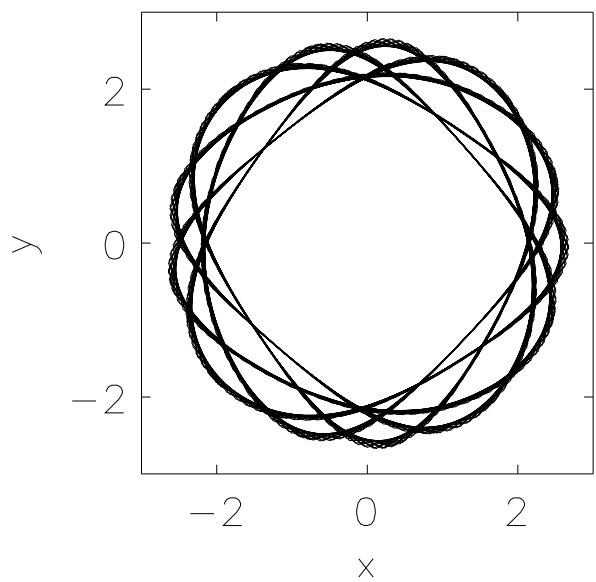
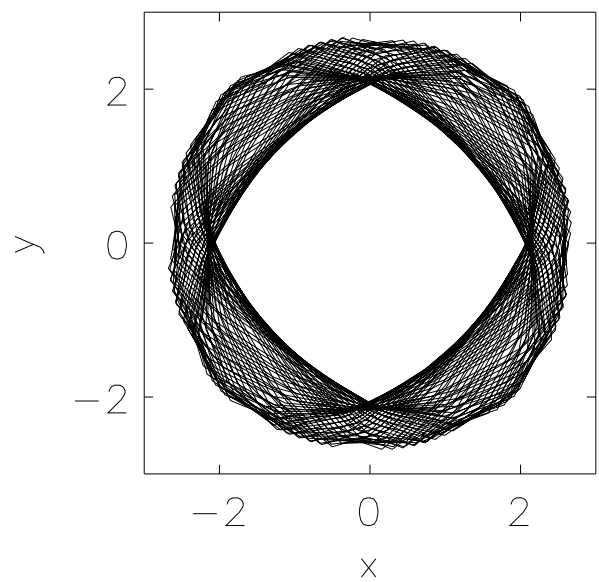
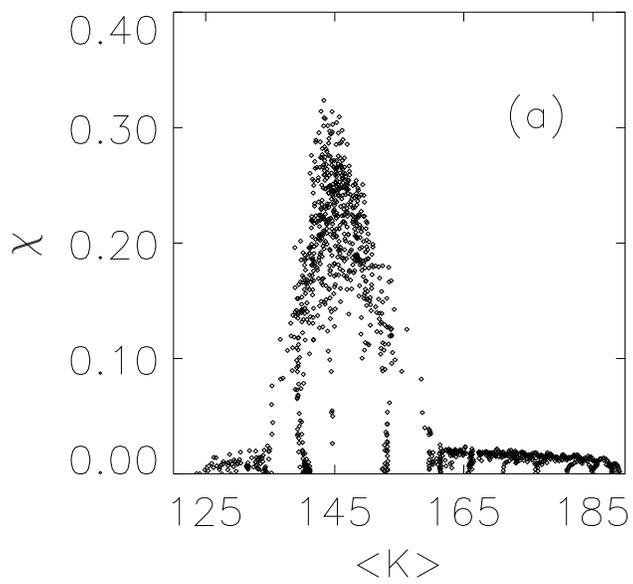
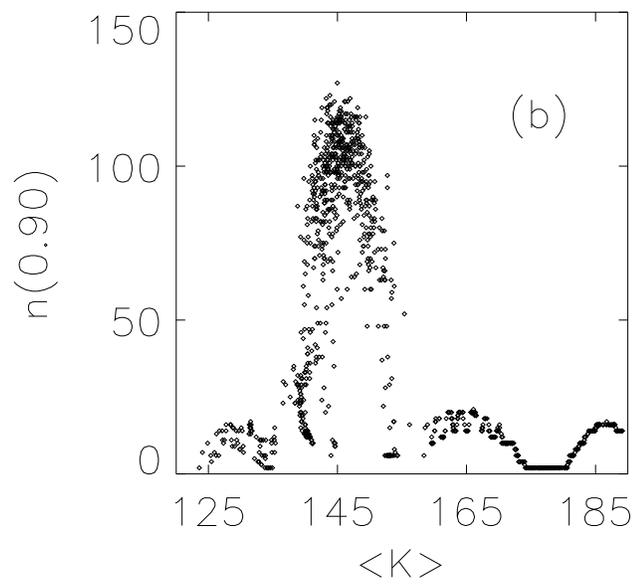
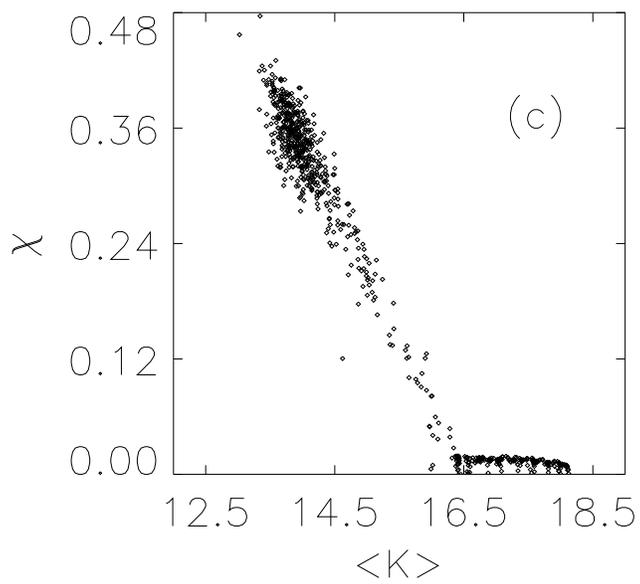
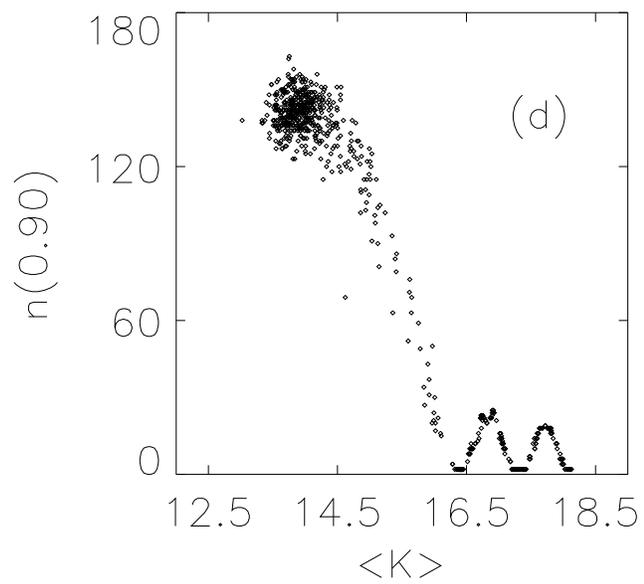
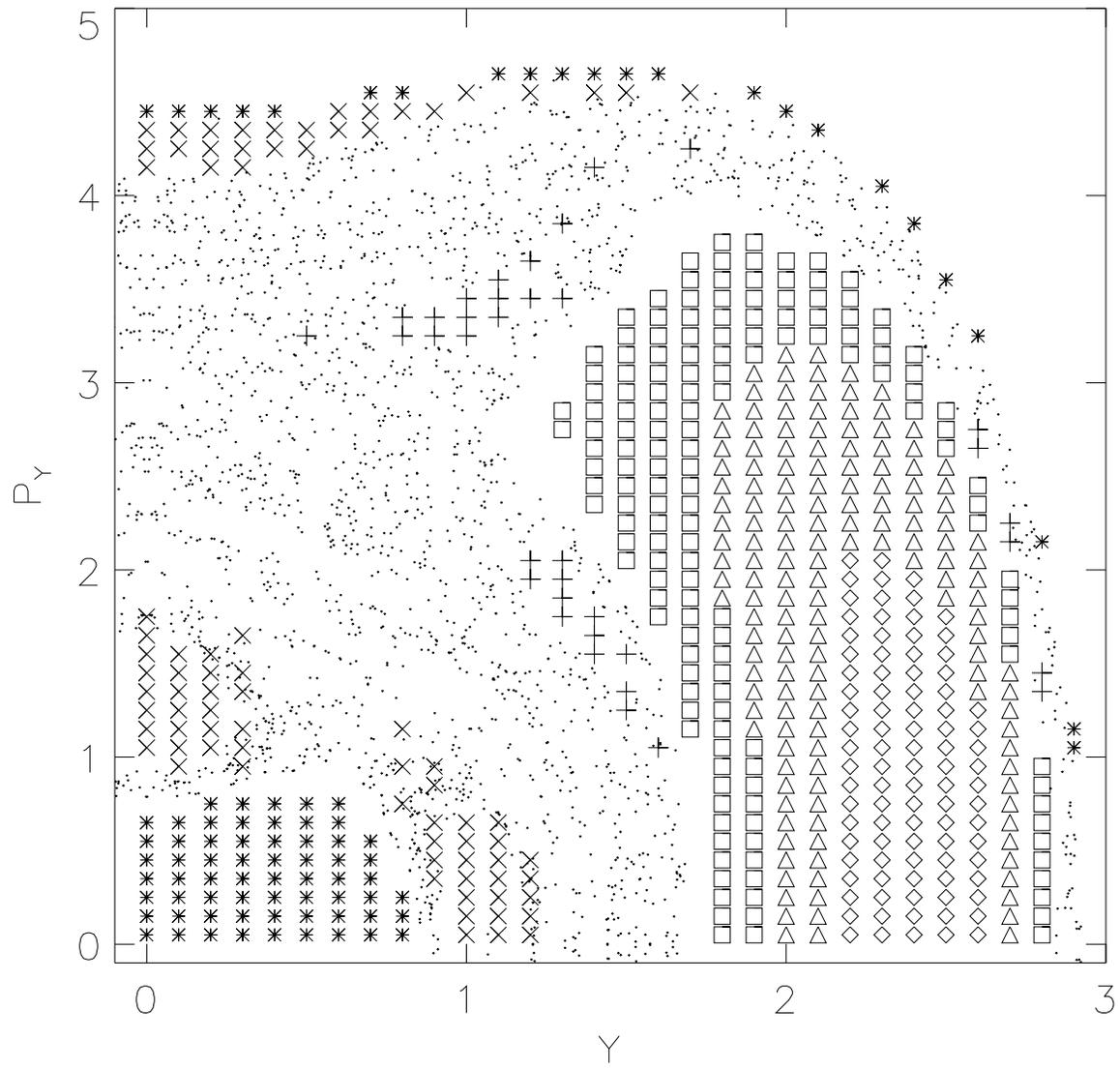
\end{document}